\documentclass[a4paper,12pt,oneside,openany]{book}
\usepackage{natbib}
\usepackage{graphics}
\usepackage{epsfig}
\usepackage{amssymb}
\usepackage[dvips]{color}
\usepackage[dvipdfm,left=35mm,right=35mm,top=30mm,bottom=30mm]{geometry}

\title{\Huge Development of binary shaped pupil mask coronagraph for the observation of exoplanets\\(Ph.D Thesis)}

\author{\LARGE Kanae HAZE} 

\date{\LARGE \today}

\begin{document}

\maketitle
\frontmatter
\tableofcontents
\mainmatter

\chapter*{Abstract}
Direct observation of extra-solar planets (exoplanets) is essential to understand 
how planetary systems were born, how they evolve, and ultimately, to identify biological signatures on these planets. 
However, the enormous contrast in flux between the central star and associated planets is the primary difficulty 
in the direct observation. 
This has required stellar coronagraphs which can improve the contrast between the star and the planet to be developed. 
Of the various kinds of coronagraphs, we focused on a binary-shaped pupil mask coronagraph. 
The reasons for using this coronagraph are that it is robust against pointing errors, 
can make observations over a wide range of wavelengths in principal and is relatively simple.  
We conducted a number of coronagraph 
experiments using a vacuum chamber and a checker-board mask, a kind of binary-shaped pupil mask, 
without active wavefront control. 
\begin{enumerate}
\item We evaluated how much the PSF subtraction contributes to 
the high contrast observation by subtracting the images obtained 
through the coronagraph. 
We improved the temperature stability by installing 
the coronagraph optics in a vacuum chamber, 
controlling the temperature of the optical bench, and covering the 
vacuum chamber with thermal insulation layers. 
A contrast of 2.3$\times 10^{-7}$ was obtained for the raw 
coronagraphic image and a contrast of 1.3$\times 10^{-9}$ 
was achieved after PSF subtraction with a He-Ne laser at 632.8nm 
wavelength. 
Thus, the contrast was improved by around two orders of magnitude from the raw contrast 
by subtracting the PSF. 
\item We also carried out multi-color/broadband experiments using Super luminescent Light Emitting Diodes (SLEDs) 
with center wavelengths of 650nm, 750nm, 800nm and 850nm in order to demonstrate 
that the binary-shaped pupil mask coronagraph makes observations over a wide range of wavelengths in principal and to demonstrate a more realistic observation. 
We achieved contrasts of 3.1$\times 10^{-7}$, 1.1$\times 10^{-6}$, 1.6$\times 10^{-6}$ and 2.5$\times 10^{-6}$ 
at the bands of 650nm, 750nm, 800nm and 850nm, respectively. 
The results show that contrast within each of the wavelength bands was significantly improved compared with non-coronagraphic optics. 
\item 
However, an existing checker-board mask with a glass substrate has the problems 
of light loss by transmission, ghosting from residual reflectance and a slightly different refractive index for each wavelength.  
Therefore, we developed a new free-standing mask with sheet metal without substrate. 
As the result of He-Ne laser experiment with the free-standing mask, 
a contrast of $1.0\times10^{-7}$ was achieved for the raw coronagraphic image by areal averaging of all of the observed dark regions. 
Speckles are the major limiting factor. 
The free-standing mask demonstrated about the same ability to improve the contrast significantly as the substrate mask. 
\end{enumerate}
We demonstrated PSF subtraction is potentially beneficial for improving contrast of a binary-shaped pupil mask 
coronagraph, 
this coronagraph produces a significant improvement in contrast 
with multi-color/broadband light sources, 
and the new free-standing mask for practical use provides superior performance of improving contrast. 
We performed the tasks necessary to make the coronagraph fit for practical use. 
In conclusion, we carried out verification test for more real coronagraphic observations. 

\chapter{Introduction}


 \section{Exoplanet}

\begin{figure}[hbp] 
\begin{center} 
\includegraphics*[width=15.5cm,angle=0]{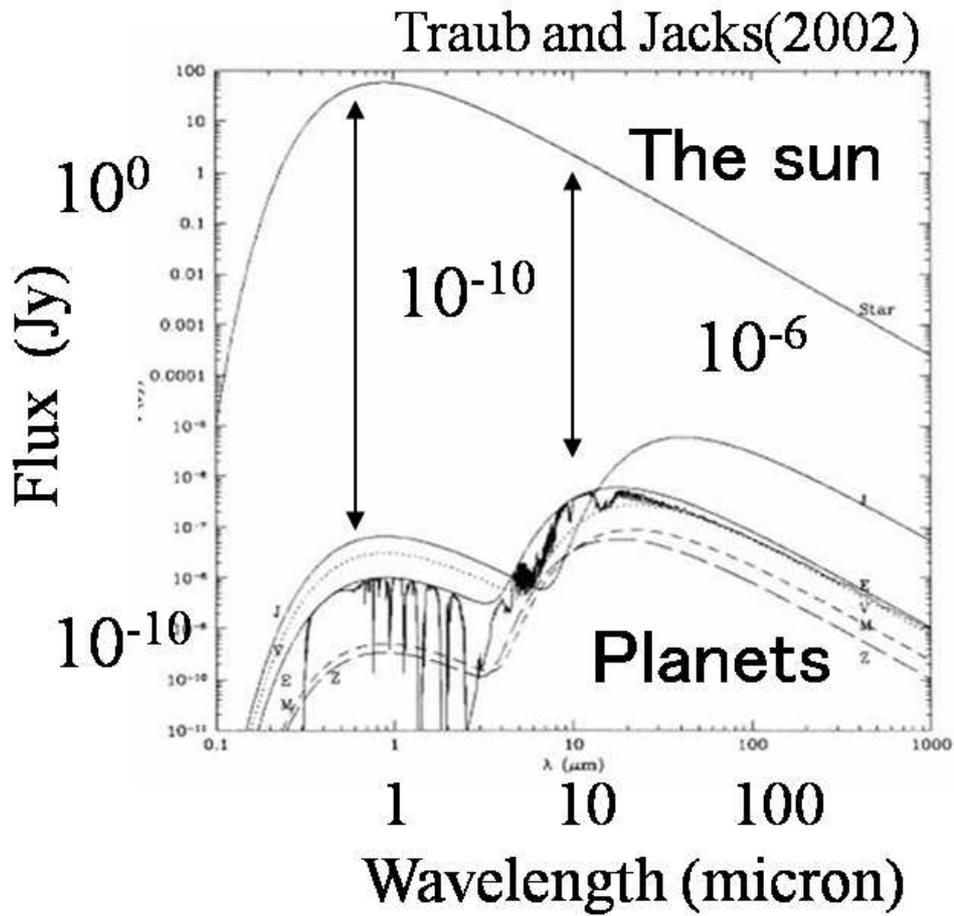} 
\end{center} 
\caption{Solar system blackbody thermal emission spectra and reflected light spectra at 10 pc, for the Sun, Jupiter (J), 
Earth (E), Venus (V), Mars (M), and zodiacal dust (Z).} 
\label{figure1} 
\end{figure}

   Since the discovery of the extra-solar planet (exoplanet), 51 Pegasi b, in 1995 \citep{Mayor}, more than 700 planets have been found. 
    A study of exoplanets is essential for understanding evolution and variety of planetary systems, and, ultimately, 
   for finding biological signatures on these planets. 
   Most of the exoplanets so far discovered are massive gas giant planets which have the planetary orbit quite unlike the Solar System. 
   We have learned that planetary system has diversity such as an eccentric planet, a Hot Jupiter and a Hot Neptune. 
   A study of exoplanets plays a complementary role vis-a-vis a study of planets in the solar system and is essential for understanding variety and time-variation of planets. 
   However, the enormous contrast in luminosity between the central star and the planet presents the primary difficulty in the direct observation of exoplanets (Fig. \ref{figure1}). 
   
 \subsection{Previously-discovered planets}
%
%
%
%

Our sun has eight planets which are Mercury, Venus, Earth, Mars, Jupiter, Saturn, Uranus and Neptune 
in order of distance from the sun. 
Mercury, Venus, Earth and Mars are terrestrial planets 
which have a layered structure of the core of iron and the rocky mantle. 
Jupiter and Saturn are gas-giant planets which have the core of mostly ice 
($<$ 20$M_{\oplus}$\footnote[2]{$M_{\oplus}$ is mass of the earth ($6.0\times 10^{27} g$).}) 
and the gas of H and He surrounding the core. 
Uranus and Neptune are considered consist almost entirely of ice. 
Each of the planetary orbits are near-coplanar circular orbits (eccentricity: e $<$ 0.1).

On the other hand, most of exoplanets so far discovered are massive gas giant planets 
which has very small radius of orbit such as Hot Jupiter and planets in elliptical orbits 
which have large eccentricities such as eccentric planets.
The exoplanets are different in orbital radius, in eccentricity and in planetary mass distribution from our solar planets.

\subsection{Conventional observation methods}


We want data of planets around the star of varied masses at different evolutionary stages, but we only obtain biased information.
There are undetectable planets because the conventional observation methods have limitations.
This is a big problem in study of the diversity and the evolution of planets.
The conventional observation methods and their limitations are as shown below.



         \subsubsection{Radial velocity observation}


         Since the discovery of the exoplanet, 51 Pegasi b, in 1995 \citep{Mayor}, many exoplanets are detected through radial velocity observations. 
         This method is to measure radial velocity of the star resulting from the planet makes stellar orbital motion. 
        The velocity vector of the star can be deduced from the displacement 
        in the parent star's spectral line caused by the Doppler effect.
        Additionally, we can know the orbital radius
        from the stellar mass and the period which are observable quantities.
        We can only know the planetary mass of the lower limit 
        from the orbital radius  and radial velocity.
%
Doppler shift can be observed from only the bright star which is relatively near the sun. 
This is because high-dispersion spectrum is necessary to detect Doppler shift.
Doppler shifts in stellar spectra have been measured to an accuracy of about 3m/s; 10m/s is common, 
and 1m/s may be the ultimate limit of this technique. 
By comparison, the solar velocity due to Jupiter is about 3m/s, and that due to Earth is about 0.01m/s.
However Doppler shift can't be detected from the bright star which is massive star, because of tiny variation.
 Only a period which is shorter than , or equal to, the data acquisition time can be detect.
This is because observing time over the planetary period is necessary to detect variation of radial velocity 
          \subsubsection{Transiting planet observation}

If the planet's orbital plane is seen nearly edge-on, a partial eclipse of the star by the planet may occur.
Transit method is to measure planetary extinction \citep{Charbonneau}.
Precise photometry has allowed us to infer the stellar limb darkening, the planet radius, the orbital inclination,
and therefore the planet's mass. 
Not only detection but also spectroscopic studies of some transiting exoplanets have been carried out (e.g., \cite{Deming, Tinetti, Swain}).  
Primary eclipse occurs when the planet passes in front of the star. 
Transmitted light from the planetary atmosphere can be obtained by comparison with stellar spectrum 
when the planet passes star and when it doesn't. 
While, Secondary eclipse occurs when the planet passes behind the star.
If we subtract the stellar spectra from the spectrum of the star and planet, 
we can detect infrared from the planetary atmosphere. 
For example, extinction between the Sun and the Jupiter can be detectable
by a photometric accuracy of one percent or less because the cross-section ratio is 1/100.
Transit doesn't require a large-scale instrument.
Thus, several dozen meters telescope and commercially available CCD camera will suffice.
Limiting conditions for observation are that planet's orbital plane must be seen nearly edge-on, 
and that we can only observe during the planet transits the star.
          \subsubsection{Gravitational microlensing observation}





         If a foreground object is aligned in one line with the Earth and a bright star in the background, 
         the light from the background star is bended by the foreground mass and can be detected by an increase in luminosity. 
         This is called gravitational microlensing, as the foreground object acts like a lens. 
         Additional observation is impossible because this event happens only once per object. 
         The microlensing can detect faint planetary mass objects (e.g., Earth-mass planets) which are either unbound to any host star or are in very  orbits, 
         because the lens object is detected by means of its mass and not its luminosity. 
         Gravitational Lensing method plays a complementary role vis-a-vis other methods.                          
         Thirteen exoplanets have been discovered by gravitational microlensing 
         and then two of them are about $3-6M_{\oplus}$ (\cite{Beaulieu}; \cite{Bennett}).

\subsection{Selection effects}


The vast majority of exoplanets detected so far have high masses. 
A lot of it depends on an observational selection effect: 
all detection methods are much more likely to discover massive planets.
Most planets have discovered around main sequence (F, G and K star). 
A star and B star in main sequence can't be detected with the radial velocity method, 
because the shift of massive star is very small.  
Radial velocity method can be used for detection of the shift of M star 
which is confined to the immediate vicinity of the sun, because M star is light but faint.
M star can be detected with the gravitational lensing method, but additional observation is impossible. 
Furthermore, the number of detection is small.
We can detect only short-period gas-giant planet which has small orbital radius in these methods. 
Exoplanet can be detected with the transit method only when it passes through the star and the orbit plane is edge-on.

There is much less information about exoplanets with indirect method than direct method, 
because indirect method is not to observe an exoplanet itself.
We have been directly observe only young planetary system which is far from the central star, 
because the contrast between planet and star is small and it is easy to separate planetary light and stellar light.


\subsection{Problems of previous exoplanet observation}
The previous observation revealed existence of many exoplanets which are far different from the solar system.
However, it is still far from more detailed and systematic observation. 
There are roughly two problems. 
  \begin{itemize}
\item One of the problem is a lack of  information of a planet itself. 
 We have little information obtained by light from a planet itself (i.e. spectroscopic data) because most of the previous observations are indirect observations.  
 \item The other is an observational selection effect. 
 Most planets are short-period gas-giant planet which have discovered around main sequence (F, G and K star). 
 \end{itemize}
Additionally, more detailed and systematic data is also required to generalize a theory of planet formation.

\section{Direct observation of exoplanets}

\subsection{Importance of direct observation}

\begin{figure}[hbp]
         \begin{center}
         \includegraphics*[width=13cm]{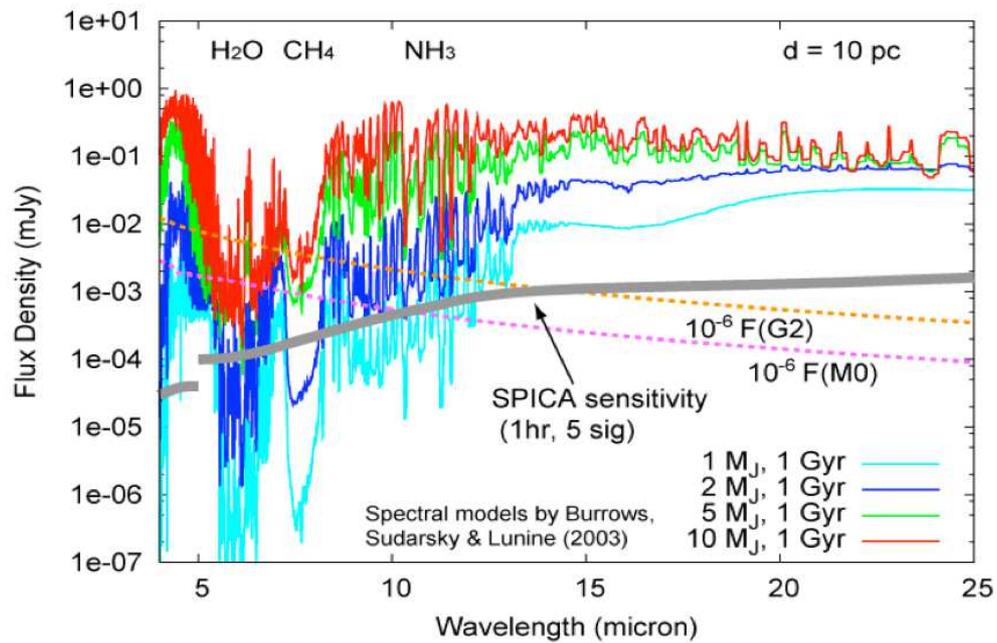}
         \end{center}
         \caption{Calculated SEDs of 1Gyrs old Jovian planets with various masses and properties relating to SPICA observations. 
         10pc is assumed as the distance to the planetary system. 
         The gray solid curve shows the sensitivity limit of imaging with SPICA. 
         The orange and purple dashed lines show the scaled SEDs of G2 and M0 type stars, respectively. This figure is from \cite{Enya2010a}.}
         \label{fig20}
         \end{figure}

Direct observations (imaging and spectroscopy) of exoplanets help understanding of their own detailed information such as size, color, luminosity and atmospheric spectrum. 
For instance, in Fig. \ref{fig20}, it is expected that the Spectral Energy Distribution (SED) of Jovian exoplanets \citep{Burrows}. 
Wavelength coverage from MIR to 3.5μm allows us to study interesting molecular features of the planetary atmospheres, e.g., H$_{2}$O, CH$_{4}$, NH$_{3}$. 
Broadband spectroscopy of exoplanets is expected to play an important role in revealing these spectral features of the planetary atmospheres.  
Also, we expect to be able to discover exoplanets that have not been discovered before 
(i.e., a planet having a larger orbital radius, a planet around AB and M type star) because of an observational selection effect. 
Therefore, direct observation is critical for understanding evolution and variety of planetary systems 
because of  obtaining the detailed information and playing a complementary role vis-a-vis other observations. 



\subsection{Difficulty of direct observation}


   Direct detection and spectroscopy of exoplanets is essential for understanding 
   how planetary systems were born, how they evolve, and, ultimately, 
   for finding biological signatures on these planets. 
   The enormous contrast in luminosity between the central star and a planet presents the primary difficulty 
   in the direct observation of exoplanets. 
   For example, if the solar system is observed from a distance, 
   the expected contrast between the central star and the planet at visible light wavelengths 
   is $\sim10^{-10}$ but is reduced to $\sim10^{-6}$ in the mid-infrared region \citep{Traub}, as shown in Fig. \ref{figure1}.    
   In this case, it is difficult to make direct observations with an ordinary telescope 
   because the planet is buried in the halo of the central star image.

   Recently, some studies on direct observations were presented \citep{Marois2008, Kalas2008}, 
   but only if they are young gas-giant planets very far from the central star. \\

\section{Coronagraph}   

\begin{figure}[hbp]
         \begin{center}
         \includegraphics*[width=13cm]{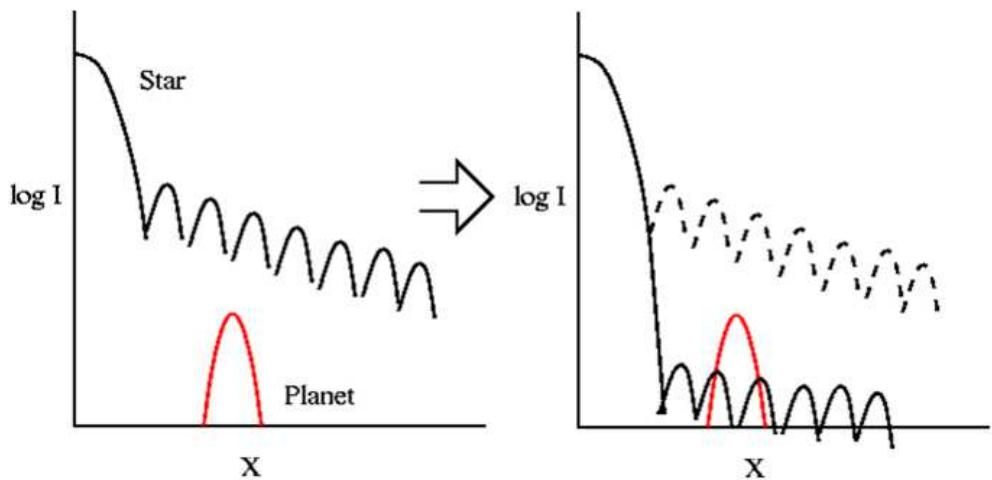}
         \end{center}
         \caption{Coronagraph can minimize the diffracted light from the central star.}
         \label{Fig14}
         \end{figure}

One of the methods the enormous contrast can be improved is a stellar coronagraph. 
Firstly, the coronagraph is evaluated for solar observation \citep{Lyot}, 
and this could provide special optics to improve contrast to minimize the diffracted light from the central star. 
It is not blocking off the light, but control of Point Spread Function (PSF), as shown in Figure \ref{Fig14}. 
The coronagraph can change the PSF and reduce the gap of luminosity between an exoplanet and its central star. 
We expect to detect the planet beyond the past observable limit and obtain the planetary own informations, 
which are color, shape, size, atmospheric spectrum and so on, by developing the stellar coronagraph. 
There are various kinds of coronagraphs, 
such as Lyot-type Coronagraph\citep{Lyot}, Phase shifting Mask Coronagraph\citep{Roddier, Rouan} , 
Pupil-plane Mask Coronagraph \citep{Jacquinot} and Phase Induced Amplitude Apodization (PIAA) Coronagraph\citep{Guyon2003}. \\

   Demonstration experiments are necessary because theory is ahead of experiments and high accuracy is required to make a theory more concrete. 
  The development race was being conducted around the world (e.g., \cite{Guyon2006}, see their Table1). 
  In this section, we describe the concepts of some coronagraphs and ongoing laboratory testing for space-telescopes. 



\begin{description}
 \item[Lyot-type Coronagraph]　
As shown in Fig.\ref{fig1-1}, this coronagraph can reduce the stellar flux by placing the focal-plane mask (i.e., an occulting spot) on the first focal plane and the Lyot stop, which blocks out the remaining rings of light from the central star while allowing most of the light from surrounding sources to pass through to the final image, on the next pupil plane.  \\

Improved performance over the original Lyot design can be obtained by the following methods:
for example, the Apodized Pupil Lyot Coronagraph (APLC), the band-limited coronagraph (BL), the Phase Mask coronagraph (PM), the Four Quadrant Phase Mask coronagraph (FQPM) and the Optical Vortex Coronagraph (OVC). \\

The Amplitude mask coronagraphs (APLC, BL) operate on the intensity of light in the focal plane, not the phase. 
The APLC described in \cite{Soummer2003a}, 
where the entrance pupil of a Lyot coronagraph with a hard edge focal plane occulter is optimally apodized. 
A slightly different approach, explored by Vanderbei et al. (2004, see their Fig.3), is to apodize the pupil after the hard edge focal plane occulter. 
As suggested by \cite{AimeSoummer}, the output of an APLC can be used as the input of a second stage APLC: these are the multistep APLCs. 
The BL uses a special kind of mask called a ``band-limited mask" in the focal plane. 
This mask is designed to block light and also manage diffraction effects caused by removal of the light (i.e., \cite{KuchnerTraub, Kuchner} ). \\

On the other hands, the Phase mask coronagraphs (PM, FQPM, OVC) introduce Phase shifts in the focal plane.  
The PM coronagraph \citep{Roddier} uses a circular-shifting focal plane mask, and a mild pupil amplitude apodization 
\citep{Guyon2000, Soummer2003a}. 
The FQPM \citep{Rouan} uses a focal plane mask that shifts two out of four quadrants of the image by $\pi$. 
The achromatic phase knife coronagraph \citep{Abe} is another form of 4QPM.
In the OVC \citep{Palacios, Foo, Swartzlander} and the Angular Groove Phase Mask Coronagraph (AGPMC) \citep{Mawet}, 
a focal plane vortex phase mask replaces the four-quadrant phase mask of the FQPM, thus avoiding the ``dead zones" of the 4QPM. 
In (r, $\theta$) polar coordinates, the mask phase is equal to m$\theta$, where m is the topological charge. \\

Phase Mask laboratory testing: 
 Monochromatic device performance has already been demonstrated and the manufacturing procedures are well-under control since their development. 
 Among them, the AGPMC \citep{Mawet, Foo}, the FQPM \citep{Carlotti} , and the Eight-Octant Phase Mask (EOPM) \citep{Murakami} are quite promising. 
 The multistage FQPM reduces the stellar flux over a  spectral range and it is a very good candidate to be associated with a spectrometer for future exoplanet imaging instruments in ground- and space-based observatories \citep{Baudoz2008}. The coronagraph gives an average transmission between $7\times10^{-6}$ and $4\times10^{-5}$  at each wavelength over a 20\% bandwidth (660-800 nm). \citep{Galicher}.

         \begin{figure}[hbp]
         \begin{center}
         \includegraphics*[width=15cm]{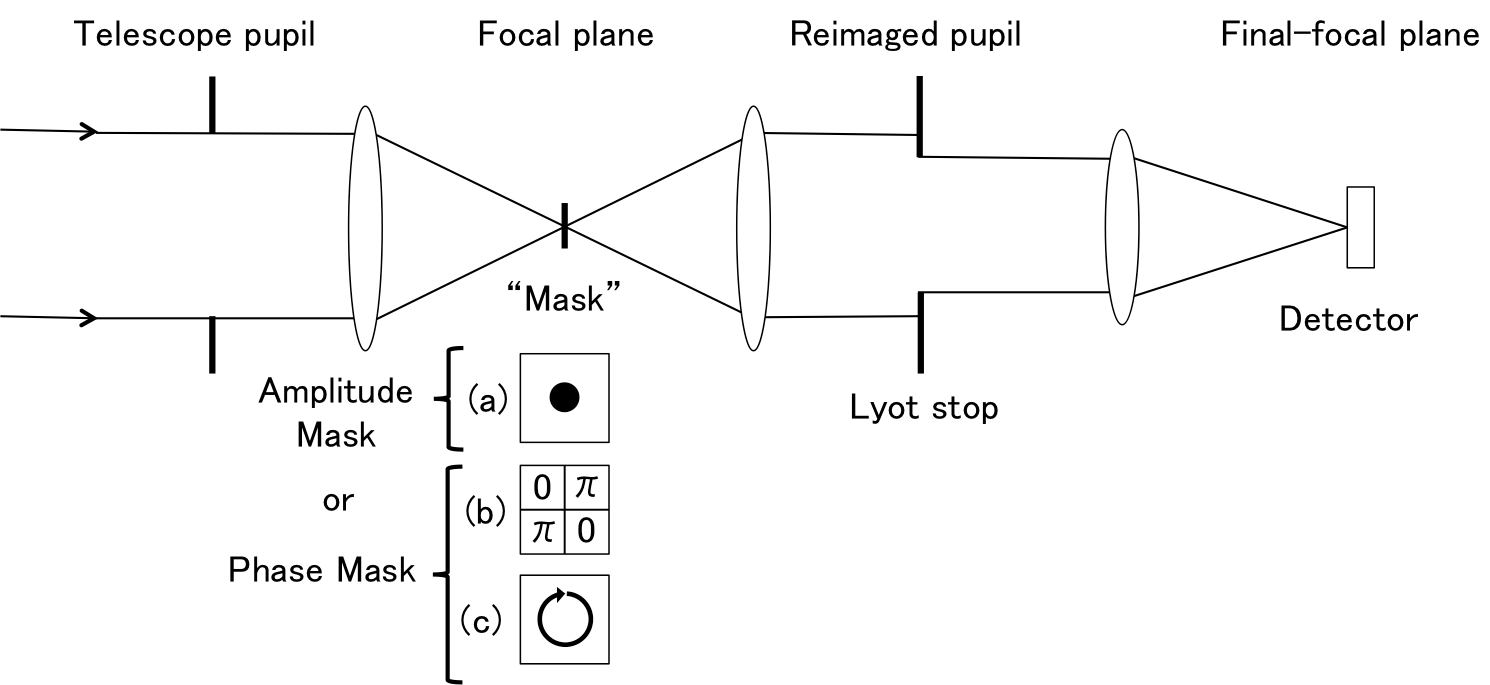}
         \end{center}
         \caption{Concept of Lyot coronagraph. 
         There are various kinds of Lyot-type coronagraph including the Amplitude mask or the Phase mask, 
         such as (a) the occulting mask, (b) the Four Quadrant Phase Mask, and (c) the optical vortex mask, on the focal plane. }
         \label{fig1-1}
         \end{figure}

 \item[Pupil-plane Mask Coronagraph] 
 The pupil complex amplitude can be modified to yield a PSF suitable for high-contrast imaging, a property used by many coronagraph concepts.
 Apodization can be performed by a pupil plane amplitude mask (Conventional Pupil Apodization, or CPA), 
 which can be continuous or binary   \citep{Jacquinot, Nisenson, Spergel, Gonsalves, Kasdin, Kasdin2005a, Kasdin2005b, Aime, Vanderbei2003, Vanderbei2003b,Vanderbei2004, Green, Tanaka, Carlotti2011, Enya2010b, Enya2011}, as shown in Fig.\ref{fig1-2}. 
Apodization by Mach-Zehnder type pupil plane interferometry was also suggested by \cite{Aime2001} to produce a continuous apodization. 
If only one-half of the focal plane is considered, an amplitude pupil apodization can be replaced by a phase-only apodization (pupil phase apodization, or PPA), 
just as phase corrections in the pupil can only cancel focal plane speckles in one-half of the field of view. 
High-contrast imaging with phase apodization was proposed by \cite{Yang}, 
who found solutions for broadband imaging and obtained contrast/throughput performances similar to amplitude apodization designs 
(although only over a quarter of the field of view). 
\cite{Codona} independently computed a PPA solution for the Hubble Space Telescope pupil to suppress diffraction in half of the field of view. \\

Pupil-plane Mask Coronagraph laboratory testing: \cite{Belikov} achieved the contrast of $4\times 10^{-8}$ using visible laser, and $\sim10^{-7}$ contrast using broadband light source with speckle nulling in a small area from 4$\lambda/D$ to 9$\lambda/D$. \\

The details of the Pupil-plane Mask Coronagraph are discussed in Chapter 1.4. 

  \begin{figure}[hbp]
\begin{center}
\includegraphics*[width=8cm,angle=0]{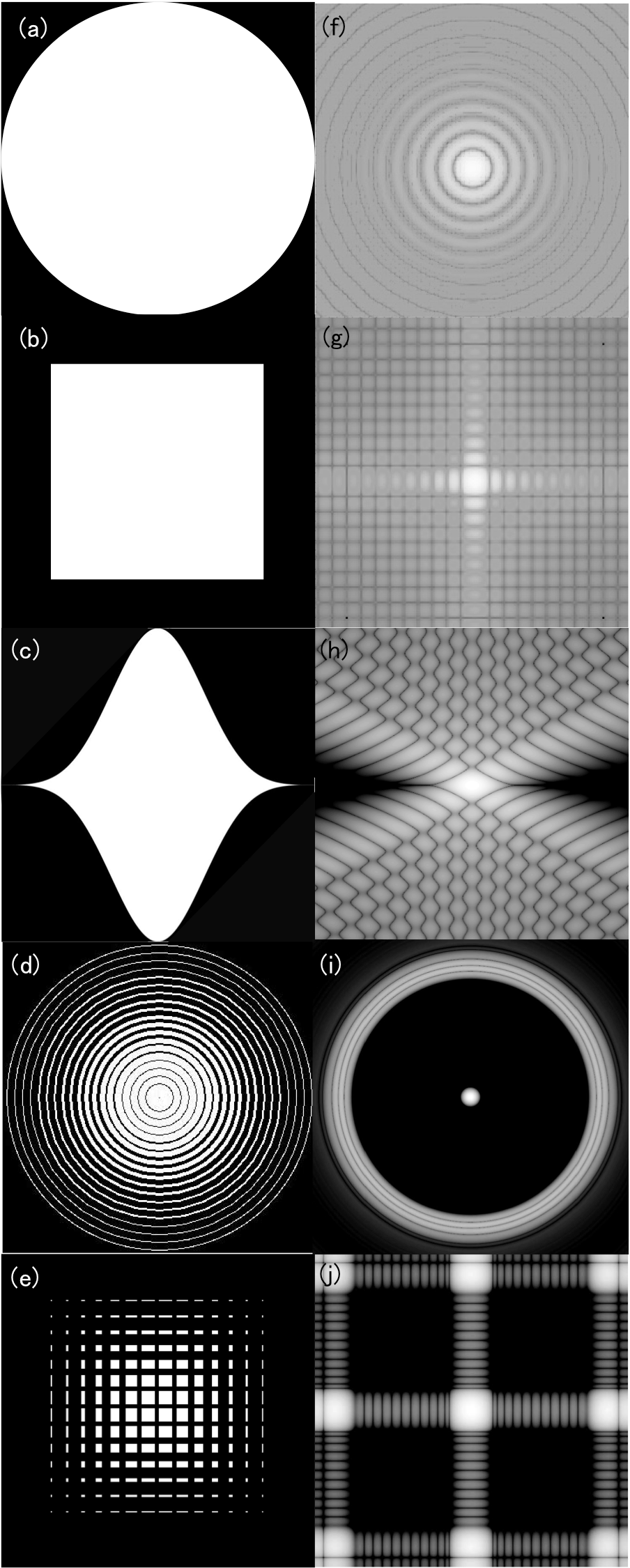}
\end{center}
\caption{Panels (a), (b), (c), (d) and (e) show the binary-shaped pupil mask designs. 
(a) is the circular aperture, (b) is the rectangular aperture, (c) is the single prolate spheroidalwave function shaped-pupil mask,  
(d) is the ring shaped pupil mask and (e) is the checkerboard pupil mask. 
The transmission through the black and white regions is 0 and 1, respectively. 
Panels (f), (g), (h), (i) and (j) show the expected (theoretical) PSFs for (a), (b), (c), (d) and (e), respectively. The gray scale is logarithmic.
Panels (c), (d), (h) and (i) from \cite{Kasdin}. Panels (e) and (j) from \cite{Vanderbei2004}. 
 }
\label{fig1-2}
\end{figure}

 \item[PIAA Coronagraph]
 As shown in Fig.\ref{fig1-3}, this coronagraph uses a lossless amplitude apodization of the pupil performed by geometric redistribution of the light rather than selective absorption 
\citep{Guyon2003, Traub2003, Guyon2005, Vanderbei2005, Martinache, Vanderbei2006, Pluzhnik}. \\
           
 PIAA laboratory testing: The laboratory experiment achieved a $2.27\times10^{-7}$ raw contrast 
   between 1.65 $\lambda$/D (inner working angle of the coronagraph configuration tested) and 4.4 $\lambda$/D (outer working angle) \citep{Guyon2010}. 
   The NASA Ames Research Center PIAA coronagraph laboratory is a highly flexible testbed operating in air \citep{Belikov2009}. 
   It is dedicated to PIAA technologies and is ideally suited to rapidly develop and validate new technologies and algorithms. 
   It uses MEMS-type deformable mirrors for wavefront control. 
   The NASA JPL High Contrast Imaging Testbed (HCIT) is a high stability vacuum testbed facility for coronagraphs. 
   PIAA is one of the coronagraph techniques tested in this lab, which provides the stable vacuum environment ultimately required to validate PIAA for flight \citep{Kern} .

         \begin{figure}[hbpt]
         \begin{center}
         \includegraphics*[width=10cm]{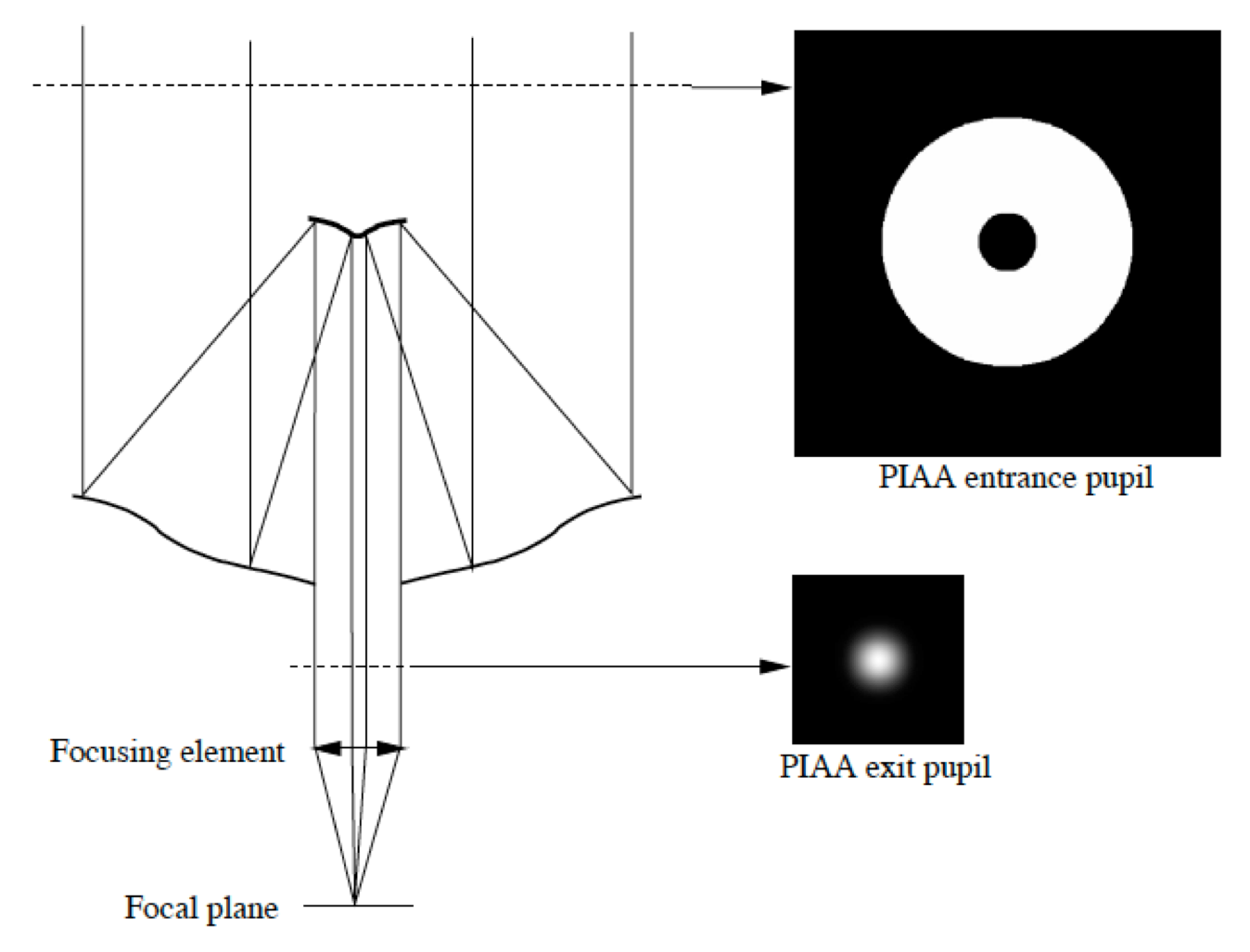}
         \end{center}
         \caption{Schematic representation of the PIAA technique from \cite{Guyon2003}. }
         \label{fig1-3}
         \end{figure}

\end{description}

There are many of the various kinds of coronagraph experiments which have a different purpose and a different demonstration performance.

\section{Binary-shaped pupil mask coronagraph}


Of the various kinds of coronagraphs, we focused on a binary-shaped pupil mask coronagraph. 
The reasons for using this coronagraph are that it is robust against pointing errors, 
can make observations over a wide range of wavelengths in principal and is relatively simple 
\citep{Jacquinot, Spergel, Vanderbei2003, Vanderbei2003b, Vanderbei2004, Kasdin, Kasdin2005a, Kasdin2005b, Green, Tanaka, Enya2010b, Enya2011}. \\

 First, we refer to the principle of pupil apodizations and binary-shaped pupil masks coronagraph. 

In this paper, we assume that telescope optics follows the Fraunhofer approximation. 
Hence, given a pupil-plane apodization function $0\leq A(x, y)\leq 1$, 
the image-plane electric field corresponding to an on-axis point source is given by the two dimensional
Fourier transform of the apodization function:

\begin{equation}
E(\xi,\zeta)=\int\int e^{2\pi i(\xi x+\zeta y)}A(x,y)dxdy,  
\end{equation}

 where $E(\xi,\zeta)$ is the magnitude of the electric field, $\xi$ and $\zeta$ are the coordinates on the image plane, x and y are the coordinates on the pupil plane, 
 and A(x,y) is the apodization function. 
 If the apodization function takes only the values zero and one, then the function represents a ``binary-shaped pupil mask''. 
 The intensity in the image plane is the square of the magnitude of the electric field.
 This intensity function is called the PSF.
Certain performance metrics guide our choice of the best pupil apodization or binary-shaped pupil mask. 
For comparison purpose, we review these metrics for some binary-shaped pupil masks, as shown in Fig. \ref{fig1-2}.　


There are various kinds of binary-shaped pupil masks, such as a gaussian pupil mask \citep{Spergel, Kasdin}, 
a ring shaped pupil mask \citep{Vanderbei2003,  Green}, and a checkerboard pupil mask \citep{Vanderbei2004, Tanaka}, as shown in Fig.\ref{fig1-2}. 
These coronagraphs can make the ``dark region", which is the area reduced the diffracted light from the central star. \\

 Second, we compare the Lyot-type coronagraph and the Binary-shaped pupil mask coronagraph 
    in order to describe the benefit of the Binary-shaped pupil mask coronagraph. 
    The Lyot-type coronagraph achieves high-contrast by placing the mask on the focal plane, 
    on the other hand, the Binary-shaped pupil mask coronagraph achieves high-contrast by placing the mask on the pupil plane. 
    Therefore, the Binary-shaped pupil mask coronagraph is robust against pointing errors compared with the Lyot-type coronagraph. 
    Furthermore, the Lyot-type coronagraph requires to shift the mask position with a wavelength, 
    but the Binary-shaped pupil mask coronagraph does not. 
     The Binary-shaped pupil mask coronagraph can make observations over a wide range of wavelengths. Note that the size of PSF scale with a wavelength. 
    Thus, this coronagraph has the advantages of robust against pointing errors and observable over a wide range of wavelengths in principal.  \\

 Finally, we refer to laboratory testing a checkerboard mask coronagraph.  
The checkerboard mask coronagraphs (i.e., Figs.\ref{fig1-2} (e)and(j)) achieve high contrast by controlling the shape of pupil \citep{Vanderbei2004, Tanaka}. 
 It should be noted that the principle of the barcode mask was presented by \cite{Kasdin2005a}, and the LOQO optimizer presented by \cite{Vanderbei1999} was used for optimization in these designs. 
The first demonstration experiments with the checkerboard pupil mask were carried out \citep{Enya2007a}. 
The size of the mask used for the experiments was 2mm.

%
               

%

  \begin{figure}[hbp]
\begin{center}
\includegraphics*[width=12cm,angle=0]{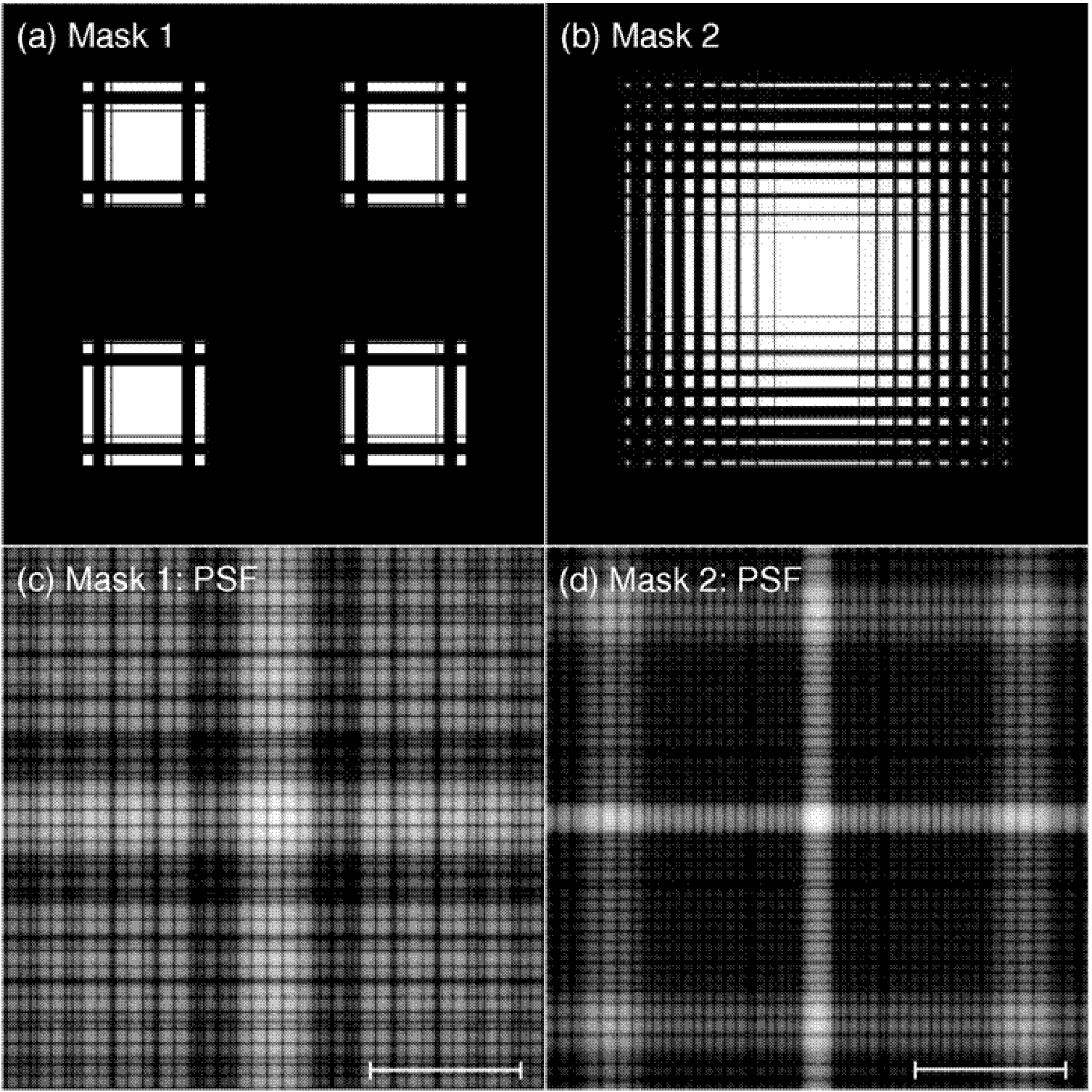}
\end{center}
\caption{Panels (a) and (b) show the Mask-1 and Mask-2 designs. The transmission through the black and white regions 
is 0 and 1, respectively. The diameter of the circumscribed circle to the transmissive part is 2mm. Panels (c) and (d) 
show the expected (theoretical) PSFs for Mask 1 and 2. The scale bar is 20$\lambda$/D. Fig.1 from \cite{Enya2007a}.}
\label{Fig1}
\end{figure}

\begin{figure}[hbp]
\begin{center}
\includegraphics*[width=13cm,angle=0]{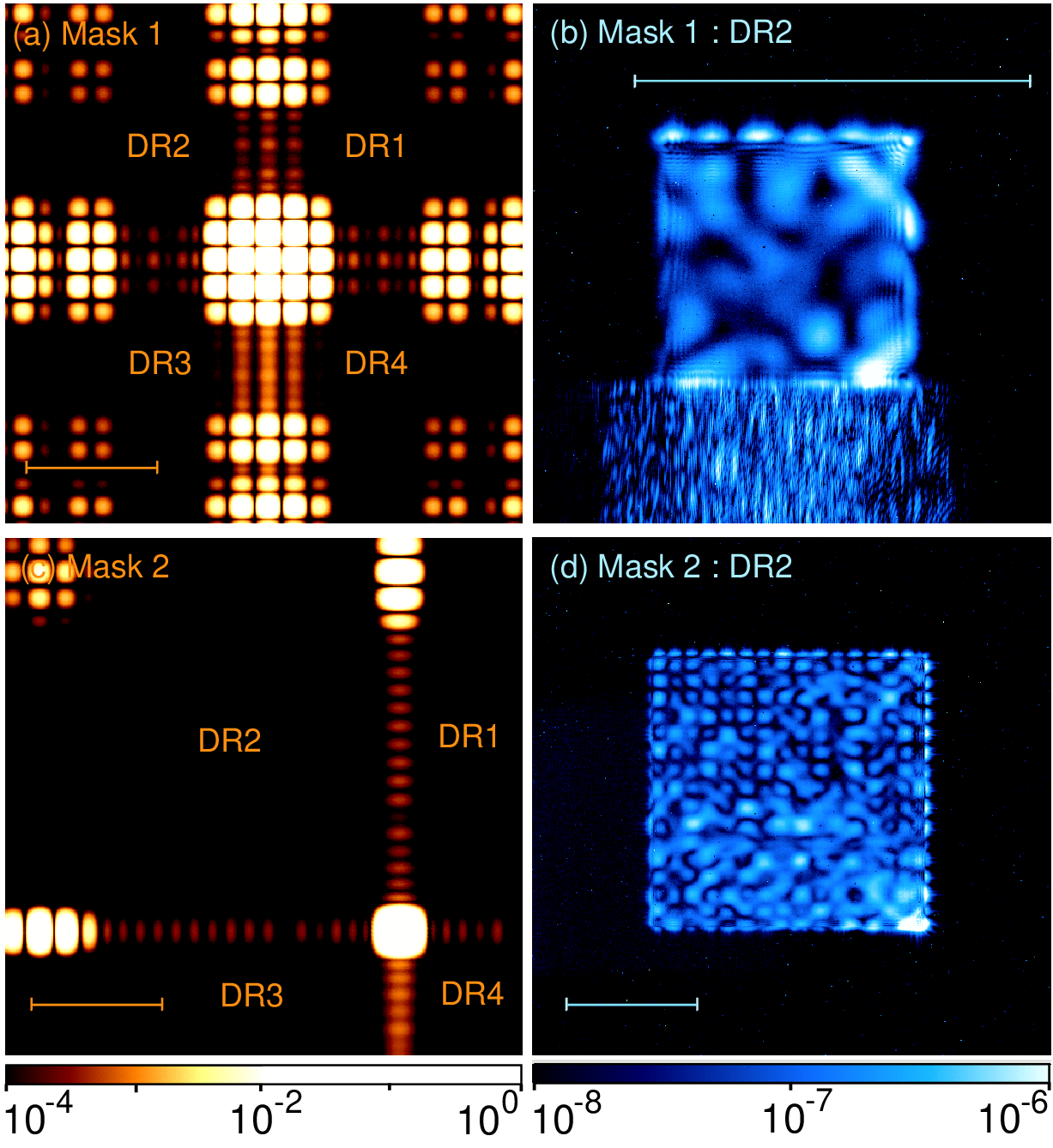}
\end{center}
\caption{Panel (a) and (c) show images including the core of the PSF for Mask-1 and Mask-2. 
The tail below the bright peak is due to a readout effect. Panels (b) and (d) are images 
of the dark region obtained with a mask with a square aperture. The prickle like pattern below 
the square aperture in (b) is the result of reflection by the support structure of the aperture for Mask-1. 
The scale bar is 10$\lambda/D$. Fig.2 from \cite{Enya2007a}.}
\label{Fig2}
\end{figure}

\cite{Enya2007a} have presented the first results from their experiments involving 
a coronagraph with a checkerboard pupil mask, as shown in Figure\ref{Fig1}, without AO at room temperature in air. 
Two masks, consisting of aluminum films on a glass substrate, were manufactured using nano-fabrication techniques 
with electron beam lithography: Mask-1 was optimized for a pupil with a 30\% central obstruction and 
Mask-2 was for a pupil without obstruction. The theoretical contrast for both masks was $10^{-7}$ and 
no AO system was employed. For both masks, the observed PSF were quite consistent 
with the theoretical ones, as shown in Figure\ref{Fig1} and  Figure\ref{Fig2}. 
Contrasts of 2.7$\times 10^{-7}$ for Mask-1 and $1.1 \times 10^{-7}$ 
for Mask-2 were achieved for raw coronagraphic images in Figure\ref{Fig2}. 
These contrasts are better than $10^{-6}$ which is the contrast between the sun and the planets at infrared wavelength. 
Brighter speckles are caused
by a combination of effects in the beam-line: wavefront errors(WFE),
multi-reflections and scattering by microscopic defects on the
surface of the optics and can therefore be reasonably considered
as being limiting factors in this experiment.

%


In the previous experiments, the high contrast performance ($<$ 10$^{-6}$) of the binary-shaped pupil mask coronagraph was confirmed. 
Further development and verification are required to install a binary-shaped pupil mask coronagraph on a telescope. 
In space telescopes, the WFE caused by imperfections in the optics is an important limiting factor in the contrast of a coronagraph. 
For an actual observation, it is necessary to make observations over a wavelength bands, and it would be beneficial to make observations using multiple bands. 
An existing checker-board mask with a glass substrate can be the problems of light loss by transmission, ghosting from residual reflectance and a slightly different refractive index for each wavelength. 
Dealing with the problems are absolutely essential for making the coronagraph fit for practical use.


\section{SPICA coronagraph}


%


There is a plan to install the stellar coronagraph in for the Space Infrared telescope for Cosmology and Astrophysics (SPICA). 
SPICA (2018 launch planned)  is an astronomical mission optimized for mid-infrared and far-infrared astronomy with a 3m class on-axis telescope cooled to $<$6K \citep{Nakagawa}.
The primary target of the SPICA coronagraph is a self-luminous Jovian extra-planets around 1-5 Gyr old G-M type stars. 
Binary shaped pupil mask coronagraph has potential for SPICA coronagraph  \citep{Enya2010a}.

SPICA coronagraph has several unique features. 
First of all, it targets not visible region but mid-infrared region. 
Mid-infrared region has a great advantage in direct observation, 
because the contrast between the sun and planets is $\sim$$10^{-6}$ at mid-infrared and $\sim$$10^{-10}$at visible (Fig. \ref{figure1}), as previously described. 
Secondly, it is the space telescope. 
Space-borne telescopes have an advantage as platforms for high contrast coronagraphs because they are free 
from air turbulence and atmospheric infrared absorption. 
Thirdly, it is the cryogenic telescope. 
The cryogenic telescope provides high sensitivity in the infrared region. 
High stability is expected as the cryogenic telescope is to be launched into deep space, the Sun-Earth L2 Halo orbit. 
In addition, the structure of the SPICA telescope, adopting a monolithic primary mirror and carefully designed secondary support, yields a clean point spread function (PSF). 
Therefore, SPICA coronagraph can be the grate opportunity for systematic exoplanet observation.

\section{Purposes of this thesis}
We set purposes of this thesis to perform the tasks necessary to make the coronagraph fit for practical use,  as described below.

\begin{enumerate}
 \item In space telescopes, the WFE caused by imperfections in the optics is an important limiting factor in the contrast of a coronagraph.  
 Subtraction of PSF is beneficial in that it removes any static WFE, and achieves  a higher contrast than the raw contrast of coronagraph (\cite{Trauger}).   
PSF subtraction is available in direct observations of exoplanets using space telescopes, which helps to improve high-contrast observations.  
We evaluate how much the PSF subtraction contributes to the high contrast observation by subtracting the images obtained through the coronagraph. 

  \item A He-Ne laser was employed as the light source in the previous experiments. 
  For an actual observation, it is necessary to make observations over a wavelength band, and it would be beneficial to make observations using multiple bands.   
 In principle the binary-shaped pupil mask coronagraph should work at all wavelengths.   
 We demonstrate this by changing the experimental system from a He-Ne laser source to broadband and multi-band light sources. 
 \item However, an existing checker-board mask with a glass substrate has the problems 
of light loss by transmission, ghosting from residual reflectance and a slightly different refractive index for each wavelength.  
Therefore, we develop a new free-standing mask with sheet metal without substrate and demonstrate the contrast performance. 
The free-standing mask is available in the infrared observation which has a grate advantage over the visible light observation in the contrast between the star and the planet.  
\end{enumerate}
These are not only essential issues to the practical use of coronagraph, but also playing important roles in more realistic demonstrations. 

A outline of this paper is as follows. 
In the next chapter "Experiments and Results", we present three kinds of experiment, concerning 1, 2, and 3 as above. 
 We describe the common part of the three experiments and then describe the experiment, result, and discussion of each of three experiments separately.  
In the third chapter "Discussions", we discuss about the issues as above and the results from a combination of each experiment.


%
%
  
%
%
%
%
%
%

\newpage

\chapter{Experiments and Results}


  %
In this section, we present three kinds of experiment, PSF subtraction experiment,  multi-color/broadband demonstrates and free-standing mask experiments. 
 We describe the common part of the three experiments and then describe each separately. 

                                             \begin{figure*}[tbp]
                                             \begin{center}
                                              \includegraphics*[width=160mm]{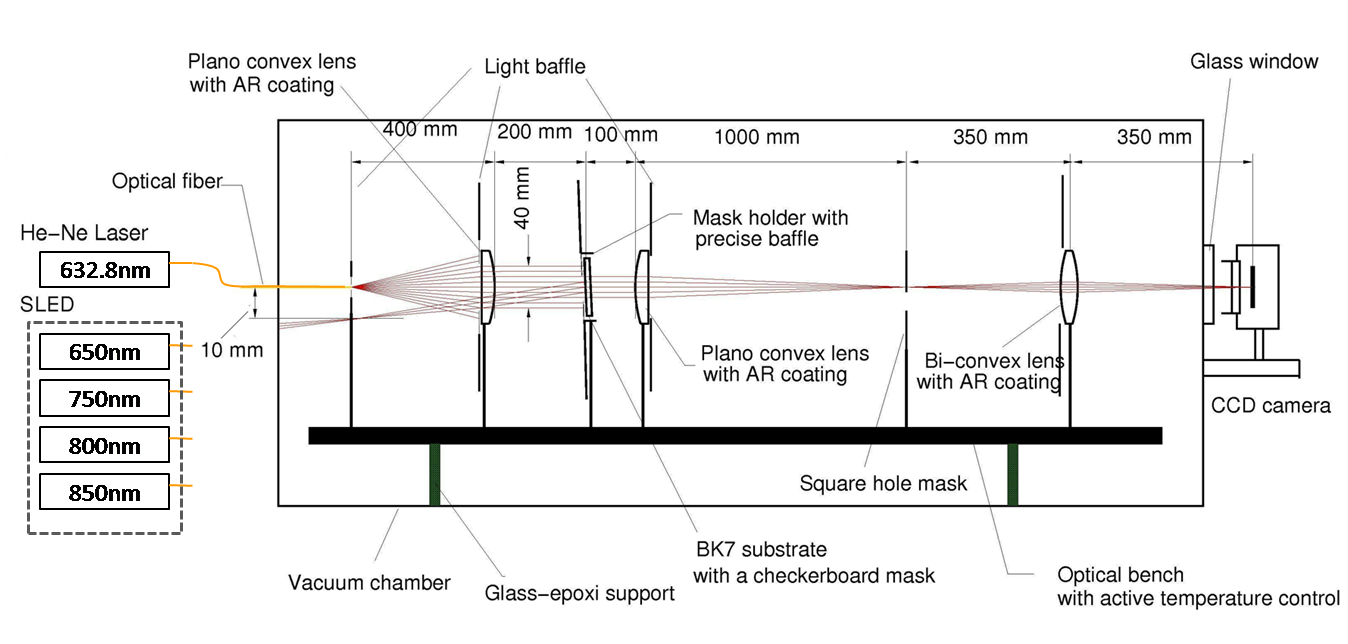}
                                             \end{center}
                                             \caption{The configuration of the experimental optics. }
                                             \label{fig2}
                                            \end{figure*}

                                     \begin{figure*}[tbp]
 　                                    \begin{center}
                                      \includegraphics*[width=120mm]{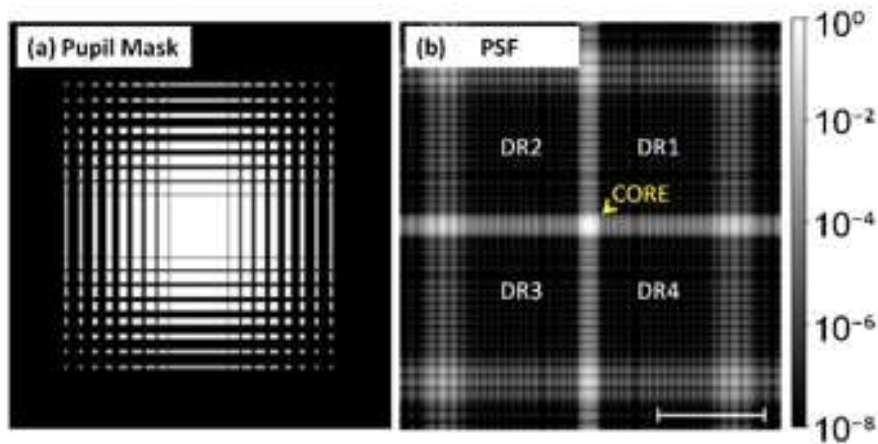}
                                      \end{center}
                                      \caption{Panel (a): the pupil mask design. 
                                      The transmissivity through the black and white regions is 0 and 1, respectively. 
                                      The diameter of the circumscribed circle to the transmissive part is 2 mm. 
                                      Panel (b): the expected (theoretical) PSF by the pupil mask. 
                                      The coronagraph can make "Dark Region (DR)", which is the area reduced the diffracted light from the central star. 
                                      There are 4 DRs nearby core of the PSF. 
                                      The scale bar corresponds to $20 \lambda /D$. 
                                      If a simulated planet bad been present in this DR, 
                                      it would have appeared as a bright spot. 
}
                                      \label{fig1}
                                     \end{figure*}
       \section{High dynamic range Optical Coronagraph Testbed (HOCT) }
        Fig.\ref{fig2} shows the experimental platform called HOCT used for this work. 
        All the experimental optics were located in a clean-room 
        at the Institute of Space and Astronautical Science/ Japan Aerospace Exploration Agency (ISAS/JAXA). 
        The coronagraph optics were set up in a vacuum chamber. 
        We used a He-Ne laser with a wavelength of 632.8nm and Super luminescent Light Emitting Diodes (SLEDs) 
        with center wavelengths of 650nm, 750nm, 800nm and 850nm, as light sources (Table \ref{table1}). 
        Light is passed into the chamber through a single-mode optical fiber. 
        The entrance beam from the optical fiber is collimated by a 50mm diameter BK7 plano-convex lens (SIGMA KOKI CO., LTD.), 
        and the collimated beam passing through the pupil mask is focused by a second plano-convex lens. 
        Active wavefront control is not applied in this work. 
        A bi-convex lens is used to reproduce the image through a window in the chamber. 
        SIGMA KOKI CO., LTD. is able to routinely achieve a surface accuracy of $\lambda$/10 for the window and 2.5$\lambda$ for the lenses. 
        BMAR coating optimized for wavelengths of 400-700 $\mu$m 
        has been applied to both sides of the lens and the window to reduce reflection at the surface. 
        A commercially available cooled CCD camera (BJ-42L, BITRAN) 
        with $2048\times2048$ pixels installed outside of the chamber is used to measure the PSF. 

        To obtain a high-contrast image, we carried out the following procedure. We measured the core and the dark region, each of which have different imaging times, separately. When the dark region was measured, we obscured the light from the core with a square hole mask inserted at the first focal plane after the pupil mask. When the core was measured, we removed the square hole mask and put in two neutral density (ND) filters. The trans- mission through the ND filters is dependent on wavelength. We show the measurements of transmission at each wavelength in Table \ref{table1}.
        \section{PSF subtraction experiment}

        \subsection{PSF subtraction}
        Subtraction of PSF is beneficial in that it removes the static wavefront error (WFE) 
        and achieves a higher contrast than the raw contrast of the coronagraph \citep{Trauger}. 
        In space telescopes, the WFE caused by imperfections in the optics is 
        an important limiting factor in the contrast of a coronagraph. 
        PSF subtraction is available in the direct observation of exoplanets using space telescopes 
        and helps improve the high contrast observation. 
        We demonstrated the PSF subtraction contributes to the high contrast observation with the checkerboard mask
        in the laboratory under vary stable environment. 
        The variations of WFE in the laboratory experiment is attributed to thermal deformation of the optics. 
        To maximize the effects of the PSF subtraction, it is important to improve the thermal stability of 
        the entire coronagraph optics. 
        
        %
        %
        %
    \subsection{Temperature stability requirement}

    In a previous study \citep{Enya2007a}, It has been found that the contrast of binary-shaped pupil mask coronagraph was better than $10^{-6}$ which is the contrast between the sun and the planets at infrared wavelength. 
    On the other hand, the contrast between the sun and the planets at visible wavelength is $\sim$$10^{-10}$. 
     Among them, the contrast between the sun and the Jupiter at visible wavelength is $\sim$$10^{-9}$. 
 We see the next step as the aim of this study is to achieve the contrast of $\sim$$10^{-9}$ required to observe giant planets directly. 
  Therefore, we consider the temperature stability of the optical system required to achieve the contrast of $\sim$$10^{-9}$. 
We arranged the data from the relationship between the error in position of the images ($\alpha$ [m]) and the contrast after the PSF subtraction in Table\ref{table2}.  
If the thermal expansion of the optical bench (made of aluminum) is $\alpha$ [m] , the temperature variation $\Delta$T [K] is 
 \begin{equation}
     \Delta T = \frac{\alpha}{CTE},    
 \end{equation}    
 where CTE=2.3$\times10^{-5}$[m/K] which is the Coefficient of thermal expansion for aluminum. 
 We considered a simplified case that the error in position of the images caused by the thermal expansion of the optical bench, 
 and arranged the data from the relationship between the error in position of the images ($\alpha$ [m]) and the temperature variation ($\Delta$T [K]) in Table\ref{table2}.  
As shown in Table\ref{table2}, the temperature variation $\Delta$T has to be $\sim$0.01K in order to aim for the contrast of $\sim$$10^{-9}$.

         \begin{table}
  \caption{Contrast vs. Temperature stability}\label{table2}
  \begin{center}
    \begin{tabular}{lcc}
      \hline
Contrast & $\alpha$ & $\Delta$T   \\ 
                &  [m]        & [K]            \\
\hline
1.9$\times10^{-8}$           &  2.2$\times10^{-6}$   &   	0.10        \\  
9.5$\times10^{-9}$           &  1.1$\times10^{-6}$     & 	 0.05       \\ 
3.8$\times10^{-9}$         &  4.4$\times10^{-7}$     &  	0.02       \\ 
1.9$\times10^{-9}$         &  2.2$\times10^{-7}$     &  	0.01       \\ 
6.3$\times10^{-10}$         &  7.4$\times10^{-8}$     &  	0.003       \\ 
\hline
    \end{tabular}
  \end{center}
\end{table}

        \subsection{Stable environment in HOCT}

        To reduce the instabilities caused by air turbulence and thermal instability, 
        the coronagraph optics were installed on an optical bench set in a vacuum chamber. 
        The optical bench was set on a support consisting of glass epoxy plates 
        to prevent thermal conduction from the chamber.
        To reduce the thermal instability, active PID temperature control was applied to the optical bench using 
        a silicon diode temperature sensor and three resistance heaters. 
        These are described in our previous study \citep{Haze}. 
        In this study, we added new functions to the experimental system as follows. 
        We used multi-layer insulation (MLI), consisting of an aluminum film on a polyethylene foam mattress, 
        on the windows in the experiment room to reduce radiation from the windows. 
        In addition, we controlled the room temperature with an air conditioner and 
        installed fans to obtain thermal equilibrium through convection. 
        To reduce thermal deformation, 
        the whole external surface of the vacuum chamber was covered with eight sheets of MLI. 
        We monitored the change of temperature on the surface of the vacuum chamber and the camera stage 
        using a total of eight silicon diode temperature sensors (DT-670A1-SD, LakeShore) 
        and a temperature monitor (Model 218, LakeShore). 
        Before introducing the MLI, a cyclical change in temperature with a period of about 1000sec was observed in all eight channels 
        and the temperature variation range was about 0.1K (peak-to-peak). 
        After introducing the MLI, this cyclical change in temperature was not observed in any of the channels and 
        the temperature variation range in every channel was significantly reduced to less than 0.03K (peak-to-peak). 
        Consequently, we were able to ascertain whether the thermal stability had an effect on the PSF subtraction 
        in our experiments.

      We use Mask2 developed in \cite{Enya2007a}. 
      The mask is a checkerboard mask, which is a type of binary-shaped pupil mask \citep{Vanderbei2004}. 
      Fig.\ref{fig1} shows the design. 
      The central brightest region of the PSF is called the "core", 
      and the four regions near to the core, in which diffracted light is reduced, is called the "dark region". 
      The contrast, inner working angle (IWA), and outer working angle (OWA) are 
      $10^{-7}$ , $3\lambda/D$, and $30\lambda/D$, respectively, 
      where $\lambda$ is wavelength and 
      $D$ is the diagonal of the checkerboard mask. 
      Optimization of the mask shape was performed using the LOQO solver presented by \cite{Vanderbei1999}. 
      The pupil mask consists of a 100nm thick aluminum film on a BK7 substrate, 
      and was manufactured using nano-fabrication technology 
      at the National Institute of Advanced Industrial Science and Technology (AIST) in Japan. 
      A standard broadband multiple-layer anti-reflection (BMAR) coating optimised for use at a wavelength of 632.8nm 
      was applied to both sides of the substrate.

\subsection{Core image}

We used a He-Ne laser in this experiment. 
The core images of the coronagraphic PSF were taken with a combination of several exposure times (0.03, 0.3, 3, 10s). 
The CCD was cooled and stabilized at $271.0\pm 0.5K (1\sigma)$. 
We inserted two ND filters as previously mentioned. 
After each imaging process, the laser source was turned off and a "dark frame" 
measurement was taken with the same exposure time and the same optical density filter. 
The dark frame was subtracted from the image with the laser light on and we obtained a "raw" coronagraphic image (Fig.\ref{fig4}a). 
This result is quite consistent with that expected from theory, as shown in Fig.\ref{fig1}b. 

\subsection{Raw image of dark region}

The dark region of the coronagraphic image was observed with a 200s exposure.
A "dark frame" was taken with a 200s exposure 
and this was then subtracted from the dark region image with the laser light on. 
The CCD was at the same temperature as when the core images were observed. 
The observed dark region of the raw coronagraphic image is shown in Fig.\ref{fig4}b. 
The "Lattice pattern" consisting of dots in the "raw" dark region corresponds to the diffraction pattern caused by 
the checkerboard mask and hence it is limited by the design of the mask. 
In other words, the current experimental contrast approximately reached the design value. 
We evaluated the contrast between the areal mean of the observed dark region and the peak of the core. 
The observed dark region is the area of the image through the square hole mask. 
As a result, a raw contrast of 2.3×$10^{-7}$ was obtained. 

\subsection{PSF subtracted image of dark region}

To obtain a contrast better than the design limit, we introduced PSF subtraction. 
The dark region of the coronagraphic image was observed 
in sets of $200s \times 18$ exposures (3600s) 2 times.
18 frames in each set of dark regions were combined and as a result 
two images totaling 3600s exposure were obtained. 
We obtained the dark region of the result of PSF subtraction using these two images. 
The images after PSF subtraction are shown in Fig.\ref{fig4}c.
There are some residual patterns on the dark region compared to the background as shown in Fig.\ref{fig4} (c).  
The dark region consists of a "lattice pattern" derived from the design of the mask, 
"speckle" from systematic errors in the experiment and "random noise" including dark current and readout noise. 
The background consists of random noise. 
Thus, the residuals on the dark region after PSF subtraction are composed of the residual of background noise and the residual of coronagraphic image (lattice pattern and speckle). 
Hence, the upper limit of contrast is 5.3$\times10^{-9}$ defined as the ratio between the standard deviation of the subtracted dark region (1$\sigma_{DR}$) and the peak of the core. 
It is necessary to factor out the residual of background noise in order to estimate the residual of coronagraphic image.  
As a simple way to factor out the residual of background noise, the standard deviation (1$\sigma$) of the coronagraphic image after PSF subtraction is defined by 
$\sigma_{}=\sqrt{\sigma_{DR}^{2}-\sigma_{BG}^{2}}$, 
where $\sigma_{DR}$ is the standard deviation of the dark region 
and $\sigma_{BG}$ is the standard deviation of the background. 
We evaluated the contrast between the $\sigma$ after PSF subtraction and the peak of the core. 
A contrast of 1.3$\times 10^{-9}$ was achieved for the PSF subtracted image. 
If a brighter light source is used, this contrast could become feasible in principal.

Fig.\ref{fig5} shows one dimensional profiles of the coronagraphic images obtained by measurement. 
Scaling by exposure time and optical density allowed smooth profiles of the core to be obtained.

%
%
 We considered detection threshold in order to carry out a practical comparison of the contrast 
 before and after the PSF subtraction. After the PSF subtraction, 
 the contrast was estimated using the standard deviation of the dark region, 
 as adopted in \citet{Biller}.
As shown in Fig.\ref{fig16}, Gaussian distribution was confirmed by plotting values of the dark region 
and 1$\sigma$ cover $\sim70\%$ of values of the dark region.  
Before the PSF subtraction, the contrast was estimated using the areal mean of the dark region. 
We checked the values of the observed dark region and confirmed 
that the areal mean also covers $\sim70\%$ of values of the dark region. 
Therefore, we estimate that the contrast was improved 
by almost two orders of magnitude when compared with the raw PSF.
In other words, it indicates that the adiabatic vacuum chamber was successful in reducing the instabilities 
caused by thermal deformation of the optics reported previously \cite{Haze}. 
It should be noted that the result shows stability of our apparatus. 
The stability and understanding of it are expected to be valuable 
in more complex future experimental environment with the chamber.

%
%
%
%
%
                                            \begin{figure}[tbp]
                                             \begin{center}
                                                \includegraphics*[width=100mm]{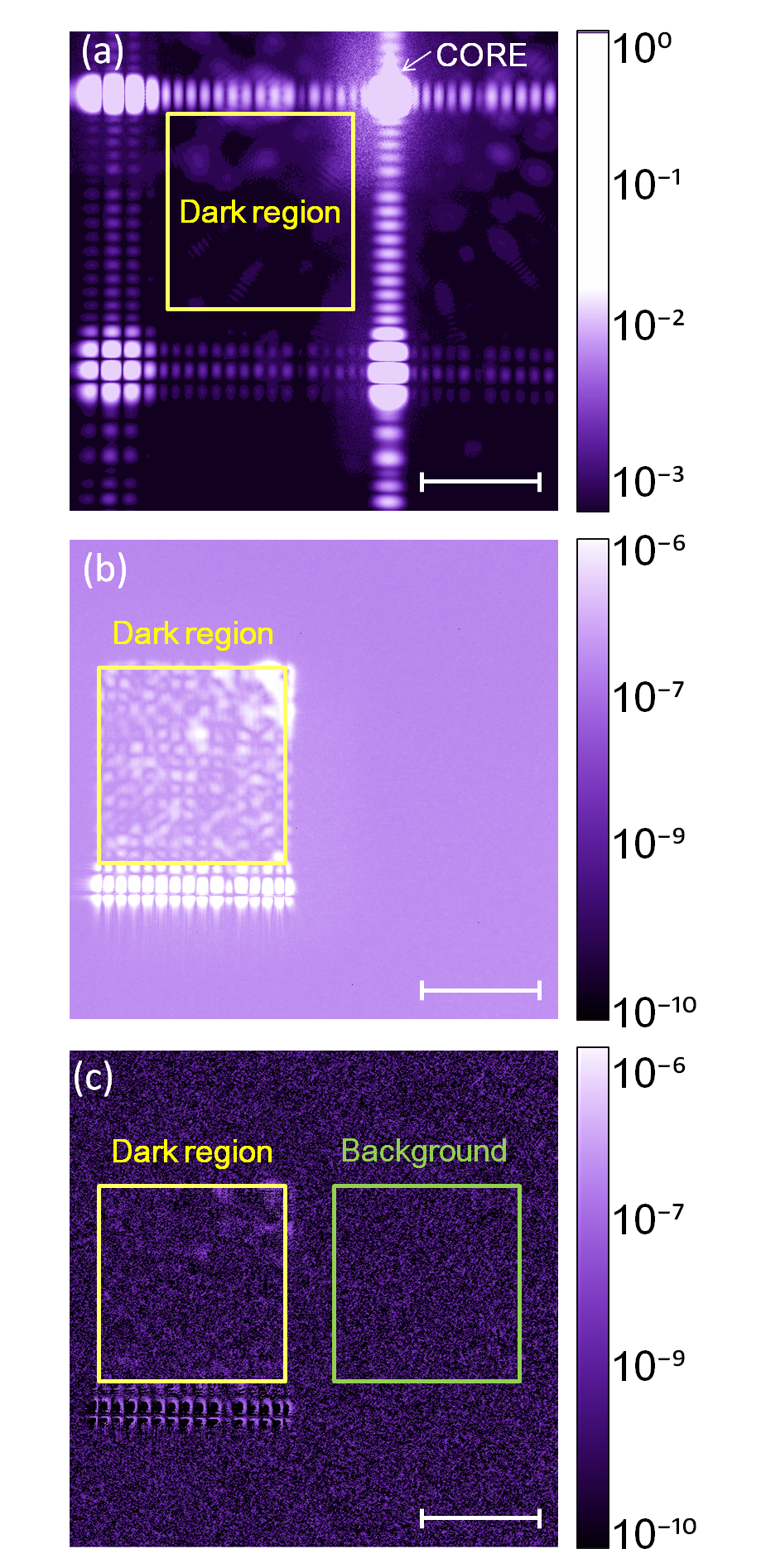}
                                             \end{center}
                                           　  \caption{Observed coronagraphic images obtained with He-Ne laser. 
                                               Panel (a): an image including the core of the PSF obtained with the ND filters. 
                                               Panel (b): a raw image of the dark region (the yellow rectangle) obtained with the square hole mask. 
                                               Panel (c): an image of the result of PSF subtraction using two raw coronagraphic images. 
                                               The green rectangle shows the area used to derive the background level in analysis. 
                                               The scale bars correspond to $10\lambda/D$.}
                                             \label{fig4}
                                            \end{figure}
　　　　　\begin{figure}[tbp]
                       \begin{center}
                          \includegraphics*[width=160mm]{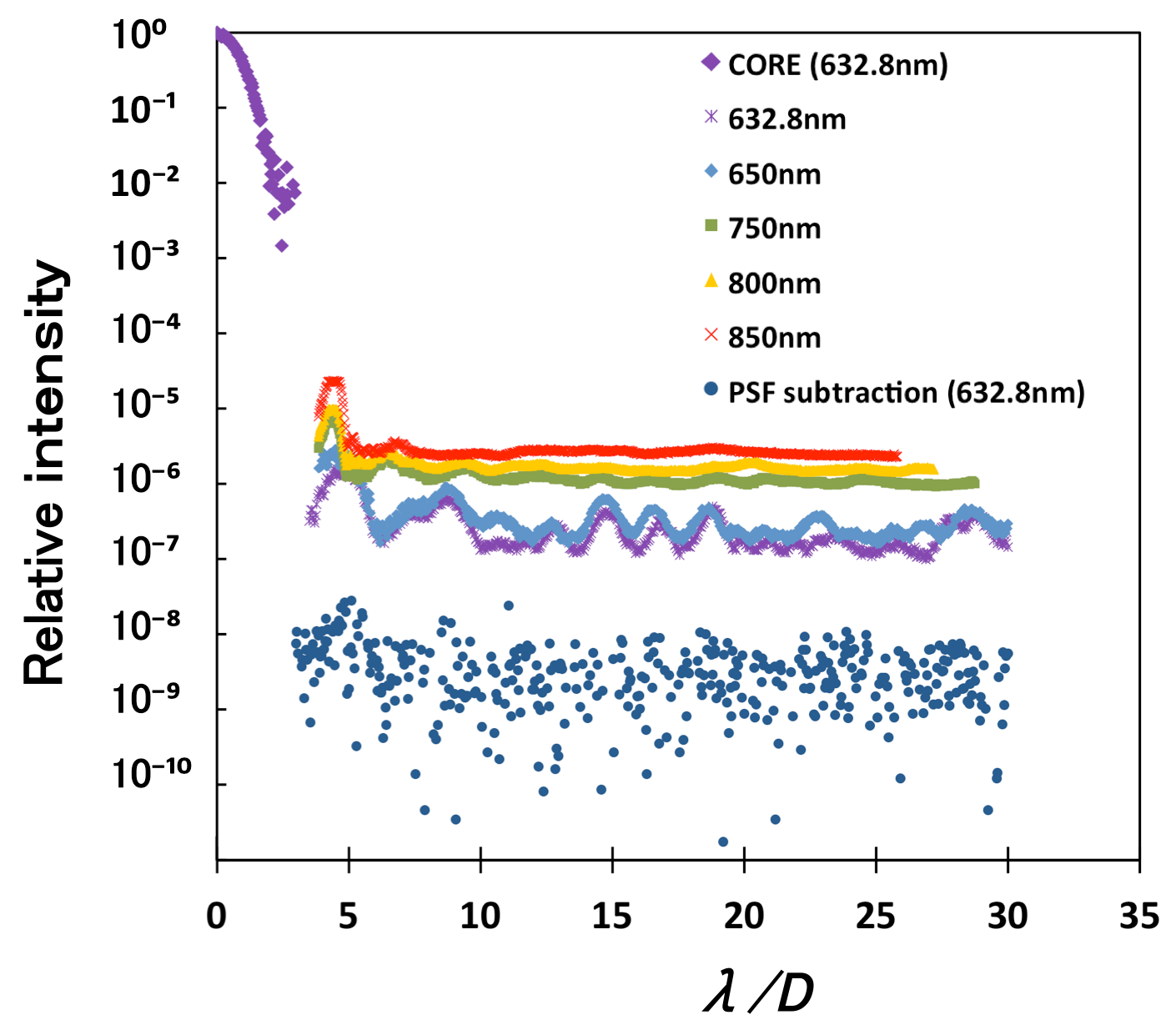}
                       \end{center}
                       \caption{Diagonal profiles of the observed coronagraphic PSF. 
                       Each profile is normalized by the peak intensity. 
                       }
                       \label{fig5}
                      \end{figure}
　　　　　\begin{figure}[tbp]
                       \begin{center}
                          \includegraphics*[width=160mm]{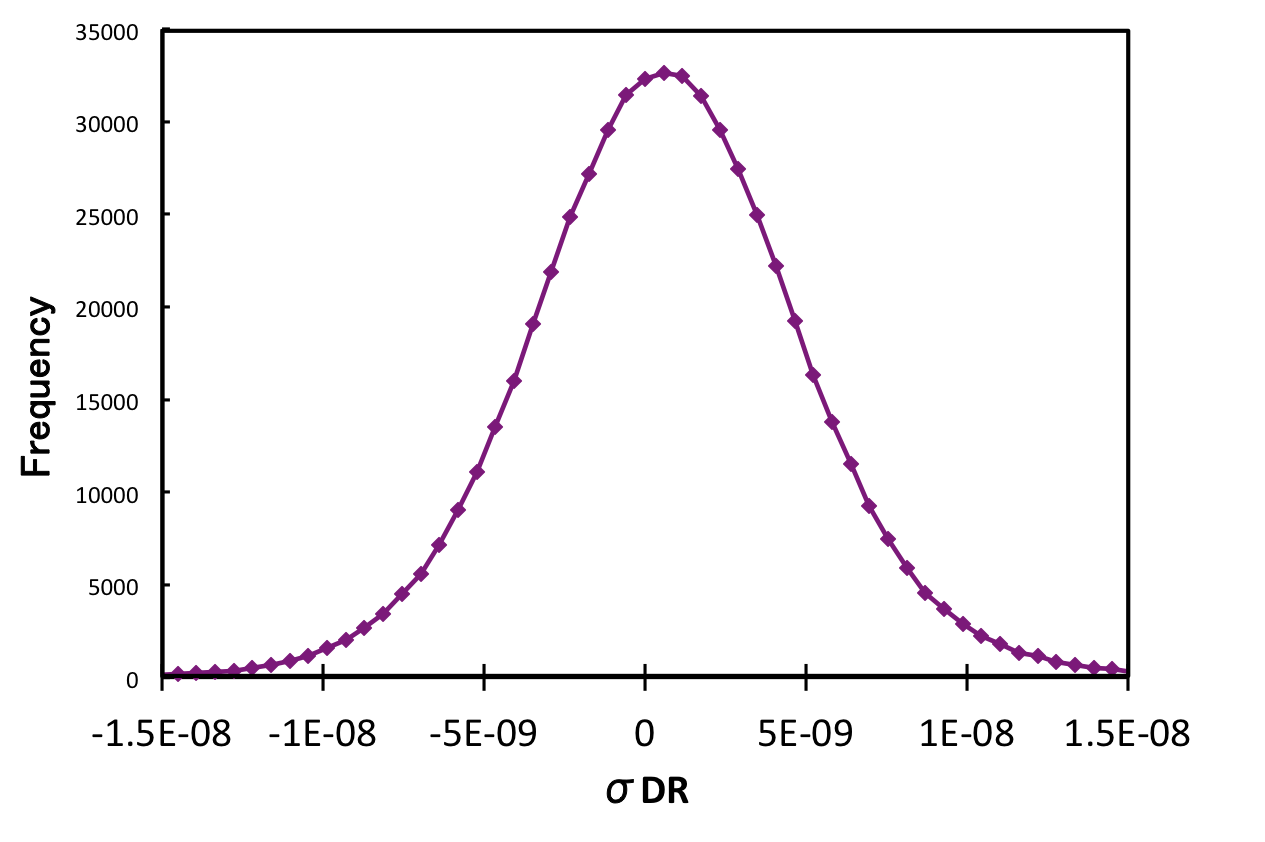}
                       \end{center}
                       \caption{Histogram of $\sigma_{DR}$. The values of $\sigma_{DR}$ normalized by the peak intensity of the core are plotted on the a axis 
                       and their frequencies on the y axis. 
                       Gaussian distribution was confirmed by plotting values of the dark region. }
                       \label{fig16}
                      \end{figure}
        \section{Multi-color/broadband experiment with SLED}
              For an actual observation, it is necessary to observe over a wavelength band and 
              it is beneficial to observe  multiple bands. 
              In principle, the binary-shaped pupil mask coronagraph can make observations over a wide range of wavelengths. 
              To demonstrate this, we used Mask2 and the light source was changed from a He-Ne laser 
              to broadband and multi-band light sources. 
              SLEDs from EXALOS 
              were used as multi-band and broadband light sources at four wavelengths. 
              The wavelengths and bandwidths are presented in Table \ref{table1}. 
              The CCD was cooled and stabilized at $271.0\pm 0.5K (1\sigma)$. 
              The temperature of CCD could not be reduced to the temperature used for the PSF subtraction experiment 
              and we assume that was because of a functional problem in the CCD cooler. 
              The wavelength was changed by connecting the optical fiber outside the chamber to each light source in turn. 
              For each of the four wavelengths, the core and dark region images were observed 
              without reconnecting the light source in order to prevent changes in intensity. 
              
              The resulting images are shown in Fig.\ref{fig3}. 
              The core images were taken with several exposure times (0.3, 0.6, 0.9, 1.2, 12, 24s) and 
              the dark region images were taken with 120s and 300s exposures. 
              As  mentioned previously, we inserted two ND filters when we took the core images.
              A "dark frame" was then subtracted from the image with the SLED light on. 
              As a result, 
              we achieved contrasts of 
              3.1$\times 10^{-7}$, 1.1$\times 10^{-6}$, 1.6$\times 10^{-6}$ and 2.5$\times 10^{-6}$ at 650nm, 750nm, 800nm and 850nm, respectively (Fig.\ref{fig5}). 
                        
              The results show that the contrast was significantly improving compared with non-coronagraphic optics within each of the wavelength bands. 
              We also found the contrast degrades as the wavelength gets longer and a ghost image was observed at longer wavelengths (Figs.\ref{fig3} b,d,f,h).
              Our suggestions on these issues are as follows.

　　　　　\begin{figure}[tbp]
                       \begin{center}
                          \includegraphics*[width=140mm]{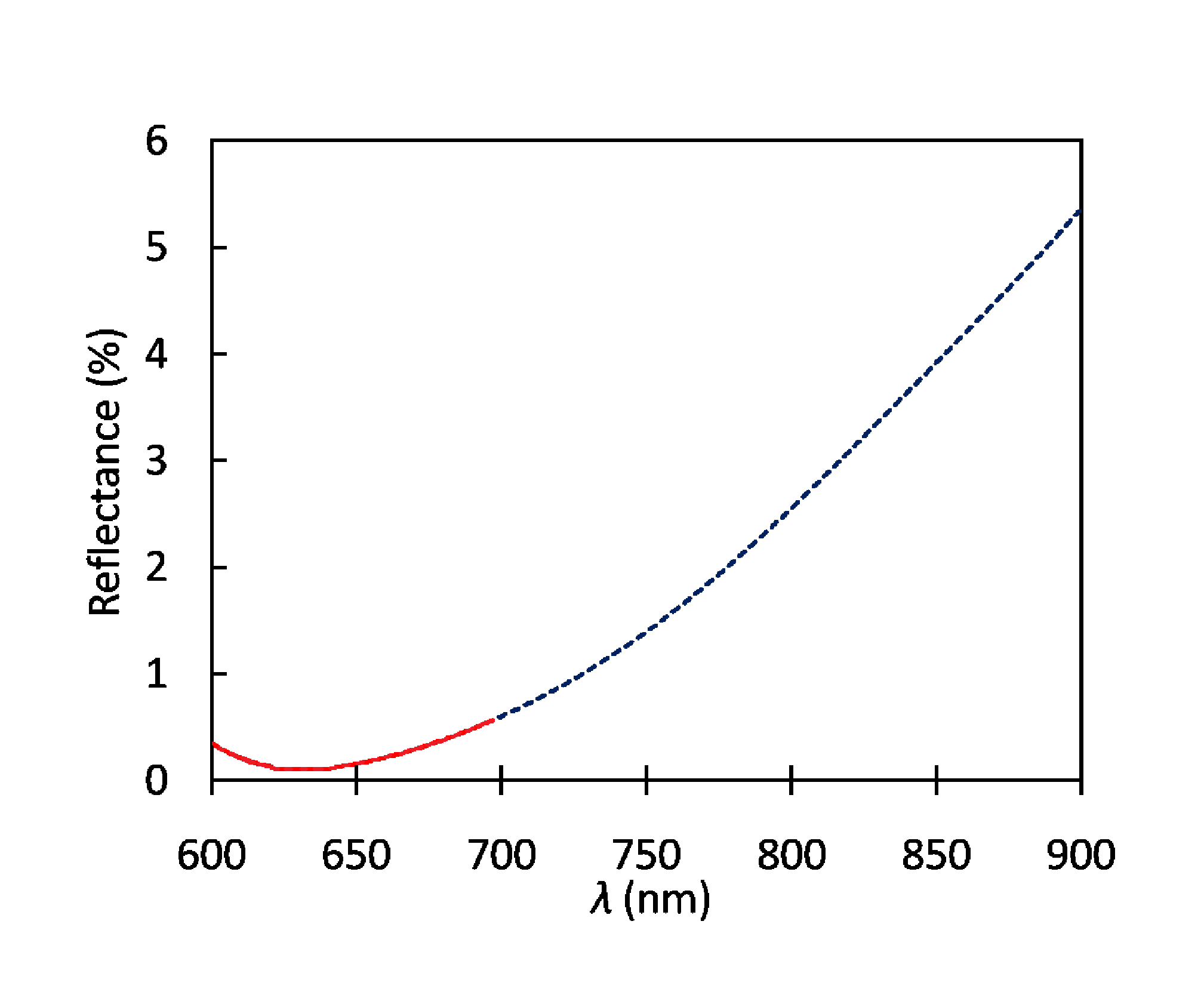}
                       \end{center}
                       \caption{Residual reflectance of BMAR coating. 
                       Red solid and blue dash lines are plots based on experiment and theoretical value, respectively.
                       }
                       \label{fig6}
                      \end{figure}

\begin{enumerate}
\item Wavelength dependency of the focal position : 
In this experiment, we changed each wavelength by reconnecting the optical fiber outside the chamber 
without changing the configuration of the optical system inside the chamber, which had been adjusted for the He-Ne laser. 
The focal planes at the various wavelengths did not coincide 
because the lenses have different refractive indices for different wavelengths of light. 
Therefore, we checked whether a major cause of the contrast deterioration 
was due to reducing the peak of the core by defocusing. 
As a result of simple ray trace analysis with ZMAX software, 
the difference of the focal plane position in the entire optical system was up to 60.6mm 
and the contrast at 850nm is deteriorated from the contrast at 632.8nm by factor of only 1.32.  
It was not a major cause of the contrast degradation.\\
\item Wavelength dependency of the residual reflectance of the BMAR coating : 
The lens we used have BMAR coat. 
The residual reflectance of the lens has a wavelength dependence, as shown in Fig.\ref{fig6}. 
The BMAR coating does not work as well at longer wavelengths. 
Ghost images may have appeared 
and the brightness in the dark region may have increased because of the increase in reflected light. 
In principle, this problem can be solved by replacing the lens with mirrors.
Therefore, we are preparing a mirror optics system, which 
requires drastic change to replace the one-dimensional optical bench with the two-dimensional testbed. 
\end{enumerate}
    %
    %
    %

                                          \begin{figure*}[tbp]
                                             \begin{center}
                                                \includegraphics*[width=140mm]{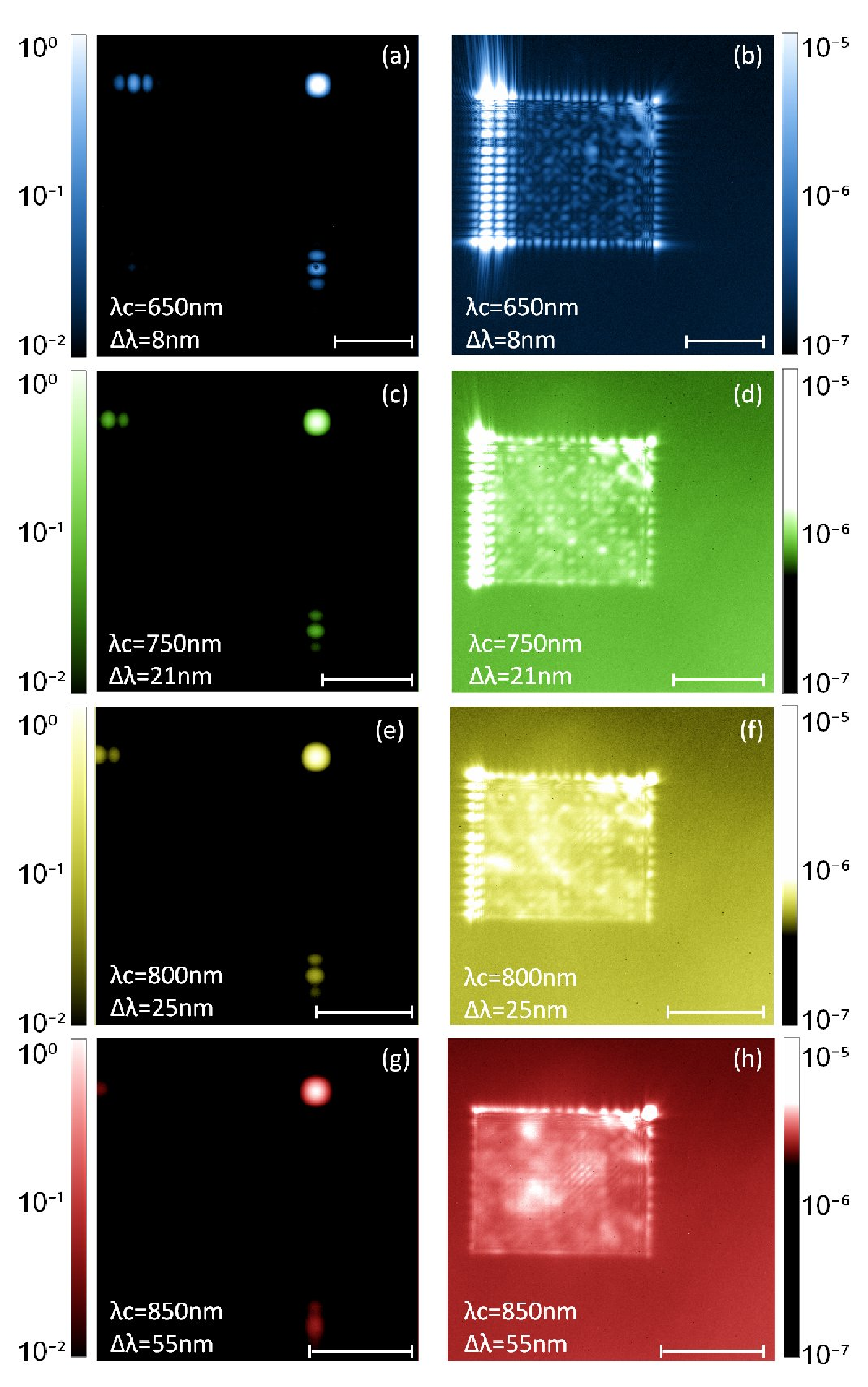}
                                             \end{center}
                                             \caption{Observed coronagraphic images obtained with SLEDs. 
                                             Panels (a), (c), (e), and (g) show images including the core of the PSF obtained with the ND filters, 
                                             and panels (b), (d), (f) and (h) show images of the dark region obtained with the square hole mask, 
                                             respectively. 
                                             The scale bars corresponds to  $10\lambda/D$. 
                                             }
                                             \label{fig3}
                                            \end{figure*}

 \begin{table}
  \caption{Property of light sources and ND filters.}\label{table1}
  \begin{center}
    \begin{tabular}{lccc}
      \hline
Light source & $\lambda_c$  & $\Delta\lambda$ & ND transmissivity \\ 
             &  [nm]        & [nm]            & [\%] \\
\hline
He-Ne laser    &  632.8   & 	-      &  	0.0088  \\  
SLED           &  650     & 	 8     &  	0.016  \\ 
SLED           &  750     & 	21     &  	0.25  \\ 
SLED           &  800     &  	25     &  	0.30  \\ 
SLED           &  850     &  	55     &  	0.32  \\ 
\hline
   \multicolumn{4}{@{}l@{}}{\hbox to 0pt{\parbox{85mm}{\footnotesize
       \par\noindent
       \footnotetext{}$*$ Total transmission through the two ND filters at $\lambda_c$.
     }\hss}}
    \end{tabular}
  \end{center}
\end{table}

               %
               %
               %

\begin{figure}[htbp]
\begin{center}
\includegraphics*[width=10cm,angle=0]{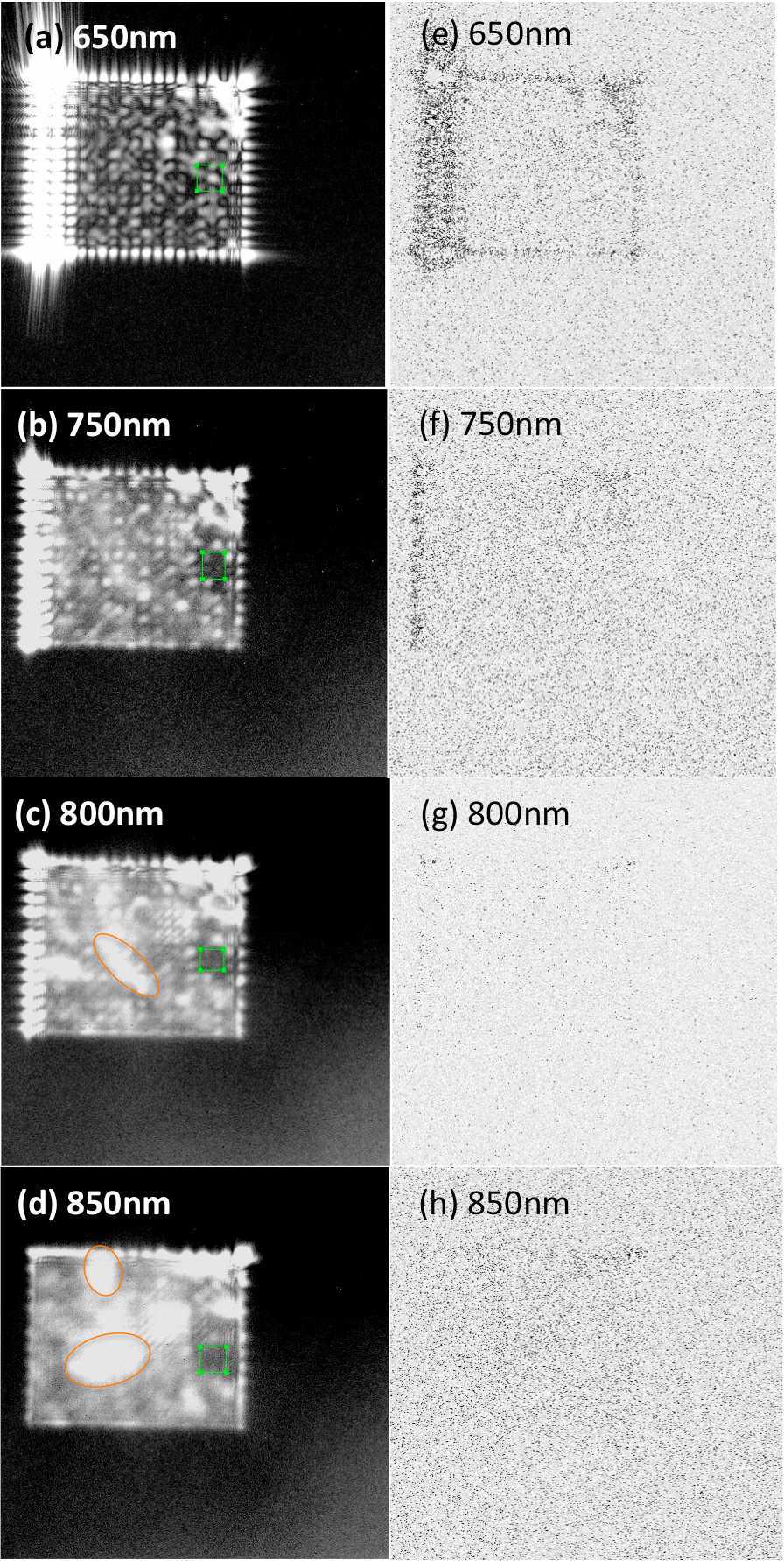}
\end{center}
\caption{Panels (a), (b), (c), and (d) show images of the dark region, respectively. The green square regions are chosen as the relatively-clean dark region.
Panels (e), (f), (g) and (h) show images of the dark region after the PSF subtraction, respectively. }
\label{Fig1-4}
\end{figure}

We checked whether the contrast of the relatively-clean dark region is achieved $10^{-7}$. 
As shown in Figs.\ref{Fig1-4} (a), (b), (c) and (d), the areal mean of the part of dark region (green square) was calculated. 
・We obtained contrasts of 2.7$\times 10^{-7}$, 9.2$\times 10^{-7}$, 1.4$\times 10^{-6}$ and 2.2$\times 10^{-6}$ at 650nm, 750nm, 800nm and 850nm, respectively. 
It was found that there are the structured ghosts (in Fig.\ref{Fig1-4}, see orange circles) and the ghosts spread all over the dark region.

 We also checked whether the ghosts are canceled by using PSF subtraction. 
 We subtracted the images were taken with 300s exposures at the same wavelength, as shown in Figs.\ref{Fig1-4} (e), (f), (g) and (h). 
As the results, we obtained contrasts of 
 2.0$\times 10^{-8}$, 3.7$\times 10^{-8}$, 2.6$\times 10^{-8}$ and 7.2$\times 10^{-8}$ at 650nm, 750nm, 800nm and 850nm, respectively. 
 It was found that the ghosts are canceled by using the PSF subtraction and the contrasts were improved.

 \section{Free-standing pupil mask experiments}
In the previous section, we have demonstrated basic properties of binary-shaped pupil mask coronagraphs. 
However, a glass substrate mask, which has been successful used in the previous experiment with visible light, has problems of opaque substrate, ghosting from residual reflectance and refractive index dependence on wavelength.  

On the other hand, a free-standing mask has advantages of no light loss by transmission, no ghosting from residual reflectance and no different refractive index for each wavelength.  
Therefore, we have developed a new free-standing mask with sheet metal without substrate. 
 We conducted free-standing mask coronagraph experiments with HOCT and evaluated contrast performance of the free-standing mask. 
Details are discussed in the following section. 

  

\subsection{New free-standing pupil mask} 

\begin{figure}[htbp]
\begin{center}
\includegraphics*[width=9cm,angle=270]{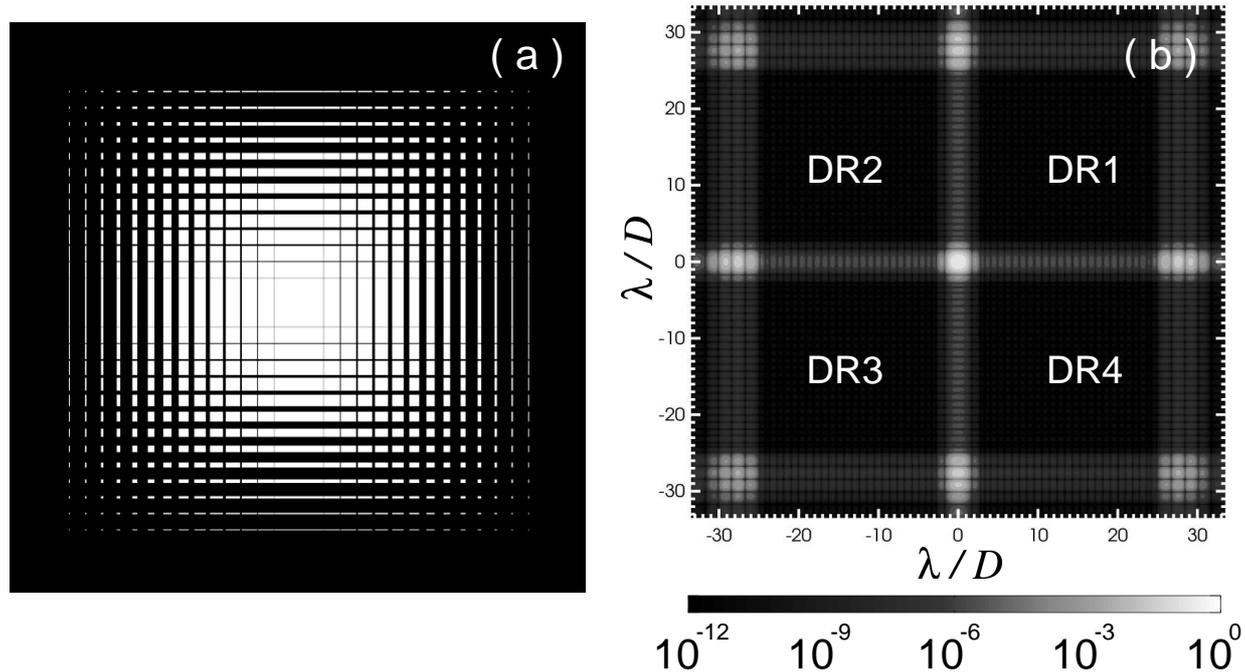}
\end{center}
\caption{Panel (a) Design of the free-standing mask. The transmissivities of the black and white regions are 0 and 1, respectively. 
Panel (b) Simulated coronagraphic PSF, using the mask design shown in panel (a). }
\label{Fig3}
\end{figure}

\begin{figure}[tbp]
                                             \begin{center}
                                                \includegraphics*[width=120mm]{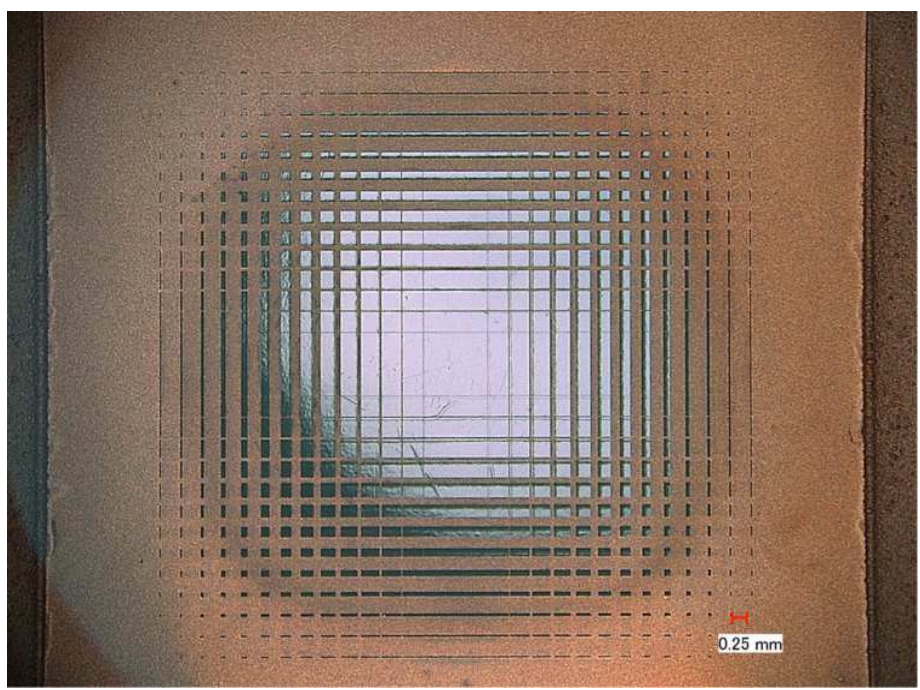}
                                             \end{center}
                                             \caption{The new free-standing mask formed of copper laminate. The transmissive region is 10mm on a side. }
                                             \label{fig7}
                                            \end{figure}

In this section, we describe design and performance the new free-standing mask (Fig.\ref{fig7}).  
The mask is a checkerboard mask (Fig.\ref{Fig3}a). 
There are four dark regions nearby core of the PSF, as shown in Fig.\ref{Fig3}b. 
The designed contrast, IWA, and OWA are $10^{-10}$ , $5.4\lambda/D$, and $50\lambda/D$, respectively.  
The free-standing mask formed of copper laminate was fabricated with high-accuracy electroforming. 
We increased the pupil-mask size to 10mm from 2mm because of the difficulty of making 2mm size free-standing mask. 

The photomask for the pattern of minute part was made first, and then the free-standing mask was made by photolithography using the photomask. 
We explain the manufacturing process of the photomask for the pattern of minute part and the free-standing mask in Appendix section.   

  

\subsection{Optical system}

                                            \begin{figure*}[tbp]
                                             \begin{center}
                                              \includegraphics*[width=160mm]{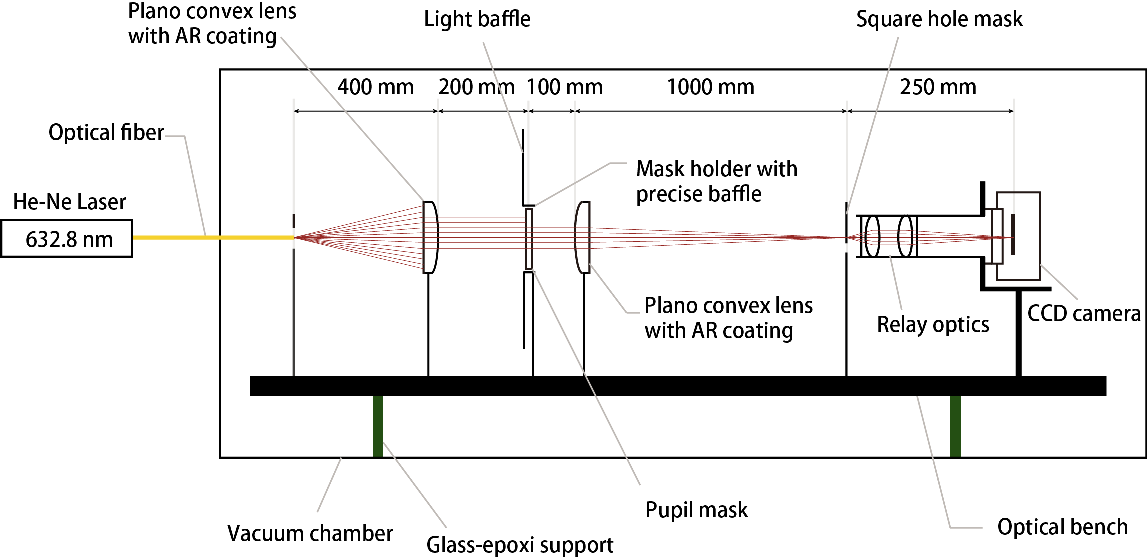}
                                             \end{center}
                                             \caption{The configuration of the experimental optics. }
                                             \label{fig10}
                                            \end{figure*}

The optical system in HOCT was shown in Fig.\ref{fig10}. 
The light source is He-Ne laser as in the PSF subtraction experiment in the Section 2.2. 
The distinctive configuration of this experiment is as follows. 
We used 10mm size of a binary-shaped pupil mask which is $10^{-10}$ of the design contrast. 
The mask is larger than 2mm size of Mask-2.  
CCD camera is set up in the chamber. 
We used 3.4$\times$ relay optics after focal plane mask with the change of the mask size and of the camera position. 


\subsection{Imaging procedure}

 To obtain a high-contrast image, we measured the core and the dark region, each of which have different exposure time, separately. 
 
 The core images of the coronagraphic PSF were taken with a combination of several sets of exposure time (0.3, 3, 10s) using two ND filters. 
 After each imaging process, the laser source was turned off and a "dark frame" 
measurement was taken with the same exposure time and the same ND filters. 
The dark frame was subtracted from the image with the laser light on and we obtained a "raw" coronagraphic image (Fig.\ref{fig11}(a)). 

 The dark region of the coronagraphic image was observed with 300s exposure using a square hole focal-plane mask. 
  We took four images (DR1 - DR4) shifting the focal plane mask,  where DR1 - DR4 are the dark regions corresponding to the quadrants around the core shown in Fig. \ref{Fig3}. 
  A "dark frame" was taken with 300s exposure and this was then subtracted from the dark region image with the laser light on. 
 The observed dark regions of the raw coronagraphic image are shown in Fig.\ref{fig11}(b)(c)(d)(e). 
 

\subsection{Contrast of the free-standing mask }

\begin{figure}[tbp]
                                             \begin{center}
                                              \includegraphics*[width=45mm]{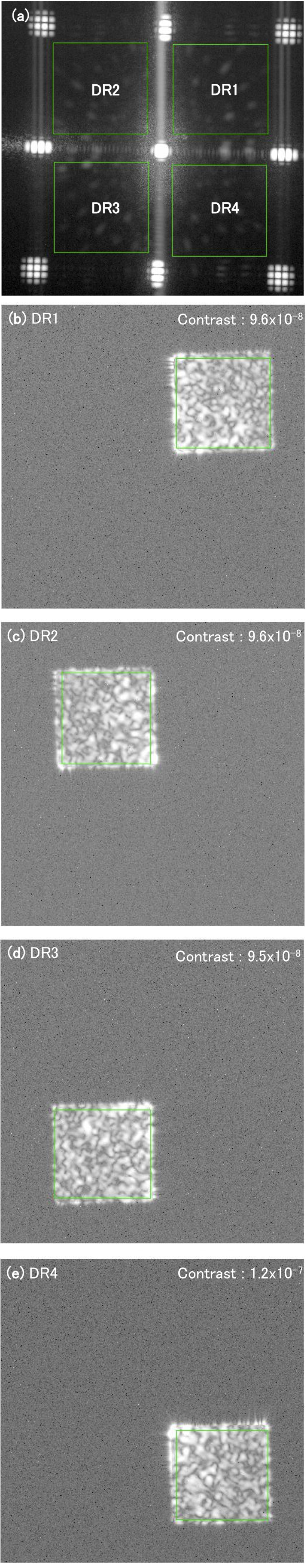}
                                             \end{center}
                                             \caption{
                                             Observed coronagraphic images before mask rotation. Panel (a): an image including the core of the PSF. 
                                             Panel (b), (c), (d) and (e): raw images of the dark regions. Each areal mean of the green rectangle in DR1 - 4 were obtained. }
                                             \label{fig11}
                                            \end{figure}

\begin{figure}[tbp]
                                             \begin{center}
                                              \includegraphics*[width=130mm]{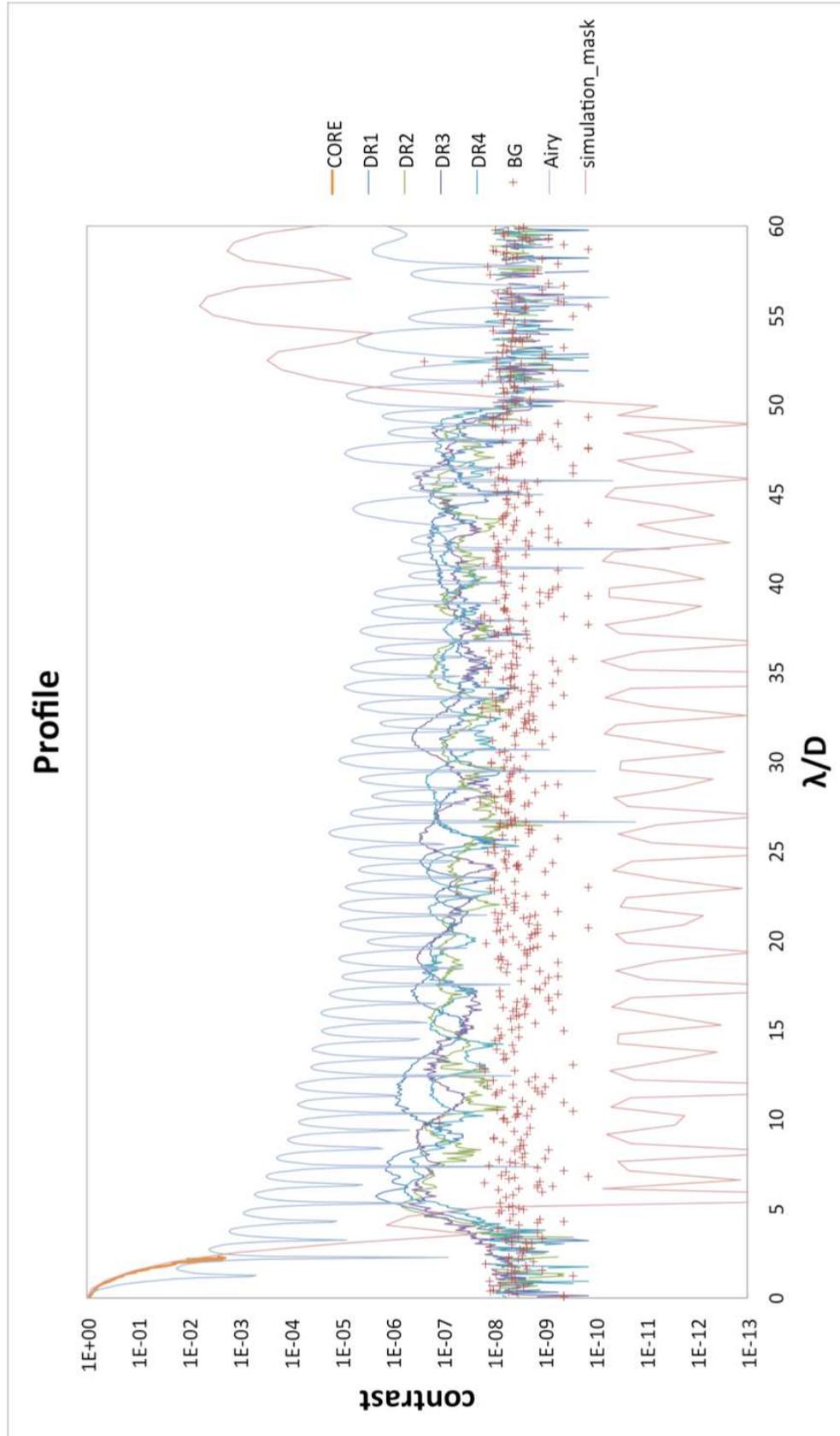}
                                             \end{center}
                                             \caption{
                                             Diagonal profiles of the observed coronagraphic PSF,  designed coronagraphic PSF and Airy PSF. 
                                             Each profile is normalized by the peak intensity.}
                                             \label{fig13}
                                            \end{figure}

Observed coronagraphic images are shown in Fig.\ref{fig11}. 
Fig.\ref{fig13} shows diagonal profiles of the observed coronagraphic PSF,  designed coronagraphic PSF and Airy PSF. 

The images and profiles of the core of the observed coronagraphic PSF are quite consistent
with those expected from theory. 
The relative intensity in most of the area of the dark region was less than $10^{-6}$, as shown in Fig.\ref{fig13} . 
We evaluated the contrast between the areal mean of the observed dark region and the peak of the core. 
The contrast values are
$1.2\times10^{-7}$, $9.5\times10^{-8}$, $9.6\times10^{-8}$ and $9.6\times10^{-8}$ for DR1, DR 2, DR 3 and DR4, respectively. 
A value of $1.0\times10^{-7}$ was obtained by averaging in the same way, but for all of the dark regions. 
Figure\ref{fig11}(b)(c)(d)(e) exhibit also irregular speckles in
the dark regions, which is not predicted by the theoretical PSF for the mask. 

One possibility for the design is that speckles are the primary limiting factor, caused by wavefront errors.

   


   　　　　　　　　　　 \begin{figure}[tbp]
                                             \begin{center}
                                              \includegraphics*[width=50mm]{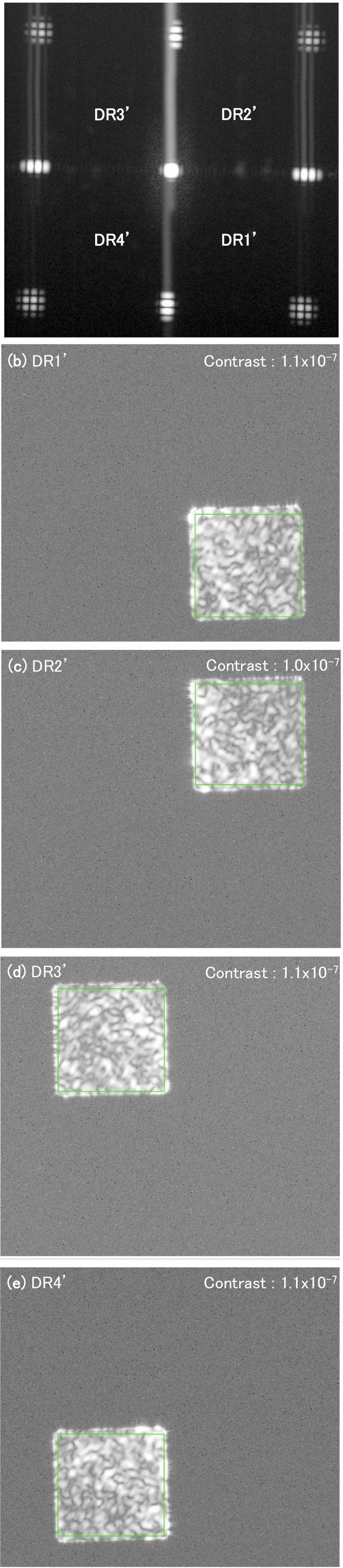}
                                             \end{center}
                                             \caption{
                                             Observed coronagraphic images after mask rotation. Panel (a): an image including the core of the PSF. 
                                             Panel (b), (c), (d) and (e): raw images of the dark regions. Each areal mean of the green rectangle in DR1' - 4' were obtained. }
                                             \label{fig12}
                                            \end{figure}

\subsection{Rotated mask subtraction}
The results of free-standing mask experiment have shown that the demonstrated contrast was at $1.0\times10^{-7}$, while the design contrast is $10^{-10}$. 
Speckles are the major limiting factor. 
We suppose that major factors of speckles are (1) pupil-mask shape error and (2) error of optical-system except the mask. 

We consider that method of rotating only the pupil-mask of the coronagraph optics will work to separate the mask shape error and the rest of error, 
because we expect that speckles from the mask shape error rotate with the mask. 

First, we subtracted the images without mask rotation for confirmation of the stability of the optical system. 
As a result, the contrast of 8.1$\times10^{-8}$ was obtained. 
The result show that the contrast was significantly improving compared with the raw contrast. 
Therefore, it is worth trying to evaluate the improvement effect of the rotated mask subtraction by using the optical system.
Then, we rotated the mask 90 degrees around a central axis perpendicular to the mask surface to confirm whether the intensity distribution of the irregular speckle pattern changed. 
After that, we subtracted the image after mask rotation (Fig.\ref{fig12}) from the image before mask rotation (Fig.\ref{fig11}). 
Rotated mask subtraction is capable of two different way of subtraction (Rotated mask subtraction 1 and 2). 

 
 \begin{enumerate} 
 \item Rotated mask subtraction 1 is the way to rotate -90 degrees the image after mask rotation and subtract it from the image before mask rotation. 
 In other words, it is a way to subtract speckles of dark regions associated with the direction of the mask. 
 If the factor of speckle is only the mask shape error, rotated mask subtraction1 is expected improving the contrast before subtraction. 
This is because speckles of dark regions are associated with the direction of the mask in this case. 
We rotated -90 degrees the images after mask rotation.  
We shifted the core image before mask rotation to fit the centers of the core before and after mask rotation, and then the shift amount was applied to dark region images. 
We increased across-the-board intensity of the dark region before mask rotation by 1.27 times to consistent the areal mean of the dark regions. 
We subtracted the images after PSF position adjustment and intensity adjustment as above. 
There are four ways to subtract (DR1-DR1', DR2-DR2', DR3-DR3', DR4-DR4' ).  
Fig.\ref{fig14} shows the PSF after rotated mask subtraction 1. 
In a comparison of speckle patterns between before and after mask rotation, the speckle patterns are completely different, as shown in Fig\ref{fig11}, \ref{fig12}. 
Also, the residuals on the rotated mask subtraction image, as shown in Fig\ref{fig14},  are different from the residuals from misalignment of images. 
Therefore, the residuals from misalignment of images are not main components of the residuals on the subtracted image. 
We evaluated the contrast between the standard deviation of the dark region after rotated mask subtraction and the peak of the core.  
The way to estimate the contrast using the standard deviation of the dark region, as adopted in \citet{Biller}, is the same way in Sec. 2.2.5.
As the result, contrasts of $1.3\times10^{-7}$, $1.4\times10^{-7}$, $1.3\times10^{-7}$ and $1.5\times10^{-7}$ at DR1 - DR1', DR2 - DR2', DR3 - DR3' and DR4 - DR4' were achieved for the images from rotated mask subtraction 1. 
The contrasts were not improving compared with the contrasts before subtraction. 
In other words, that means the factor of speckle is not only the mask shape error. 
These results show the factor of speckles include the error of optical-system except the mask.   
 We consider that the error of optical-system except the mask is composed of WFE,  uneven intensity, stray light and so on. 

 
  \item Rotated mask subtraction 2 is the way to subtract the images after from before mask rotation.
 If the mask shape and repeatability of the mask position are perfect, rotated mask subtraction 2 is expected improving the contrast before subtraction. 
This is because the error of optical-system except the mask like WFE is common to before and after mask rotation in this case. 
We shifted the core image before mask rotation to fit the centers of the core before and after mask rotation, and then the shift amount was applied to dark region images. 
We increased across-the-board intensity of the dark region before mask rotation by 1.27 times to consistent the areal mean of the dark regions. 
We subtracted the images after PSF position adjustment and intensity adjustment as above. 
There are four ways to subtract (DR1-DR2', DR2-DR3', DR3-DR4', DR4-DR1' ).  
Fig.\ref{fig15} shows the PSF after rotated mask subtraction 2. 
We evaluated the contrast between the standard deviation of the dark region after rotated mask subtraction and the peak of the core.  
As the result, contrasts of $1.2\times10^{-7}$, $1.7\times10^{-7}$, $1.7\times10^{-7}$, $1.4\times10^{-7}$ at DR1 - DR2', DR2 - DR3', DR3 - DR4', DR4 - DR1' were achieved for the images from rotated mask subtraction 2. 
The contrasts were not improving compared with the contrasts before subtraction. 
In other words, that means the factor of speckle is not only the mask shape error. 
These results show the mask shape and/or repeatability of the mask position are not perfect. 
  

\end{enumerate}  

These results of rotated mask subtraction show the factor of speckle is not only the mask shape error. 
Therefore, the factor of speckles include the error of optical-system except the mask.  
The factor of speckles is the error of optical-system except the mask (i.e., WFE, uneven intensity and stray light) or the combined effect of the mask shape error and the error of optical-system except the mask.   
The useful way to distinguish among the mask shape error, the error of optical-system except the mask and the repeatability error of the mask position is to simulate the effects of these errors. \\


\subsection{Numerical simulations of some mask shape errors }
\begin{figure}[tbp]
                                             \begin{center}
                                              \includegraphics*[width=80mm]{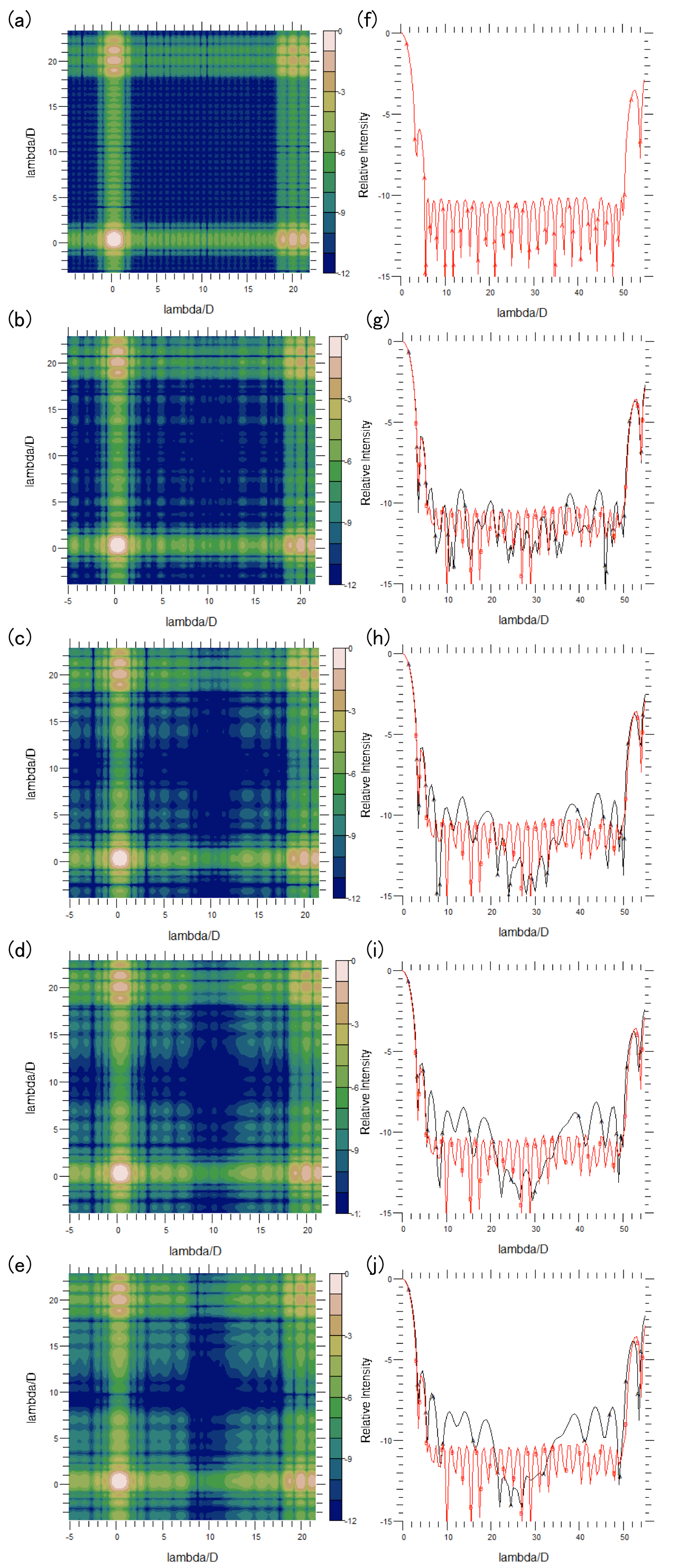}
                                             \end{center}
                                             \caption{Simulated PSFs of the free-standing mask. 
                                             (a) and (f): Simulated PSFs from the perfect mask.
                                             (b) and (g): Simulated PSFs from across-the-board 5$\mu$m increase in the line width of the mask. 
                                             (c) and (h): Simulated PSFs from across-the-board10$\mu$m increase in the line width of the mask. 
                                             (d) and (i): Simulated PSFs from across-the-board 15$\mu$m increase in the line width of the mask. 
                                             (e) and (j): Simulated PSFs from across-the-board 20$\mu$m increase in the line width of the mask. 
                                             (a), (b), (c), (d) and (e) are simulated images.
                                             (f), (g), (h), (i) and (j) are simulated profiles. 
                                             Red line shows the PSF from the perfect mask. 
                                             Black line shows the PSF from the imperfect mask. 
                                             The scale is logarithmic. }
                                             \label{Fig1-5}
                                            \end{figure}

\begin{figure}[htbp]　　　　　　　　
                                             \begin{center}
                                              \includegraphics*[width=45mm]{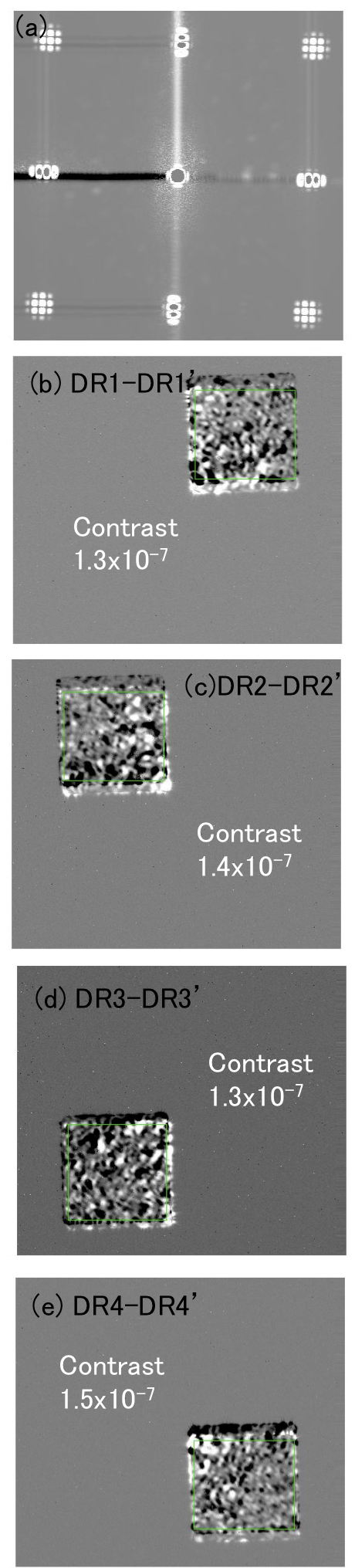}
                                             \end{center}
                                             \caption{
                                             Rotated mask subtraction 1: the way to rotate -90 degrees the image after mask rotation and subtract it from the image before mask rotation.
                                             Panel (a):  shifting the core image before mask rotation to fit the centers of the core before and after mask rotation. 
                                            Panel (b), (c), (d) and (e):  images of the results of rotated mask subtraction 1 (DR1-DR1', DR2-DR2', DR3-DR3', DR4-DR4'). }
                                             \label{fig14}
                                            \end{figure}

\begin{figure}[htbp]
                                             \begin{center}
                                              \includegraphics*[width=45mm]{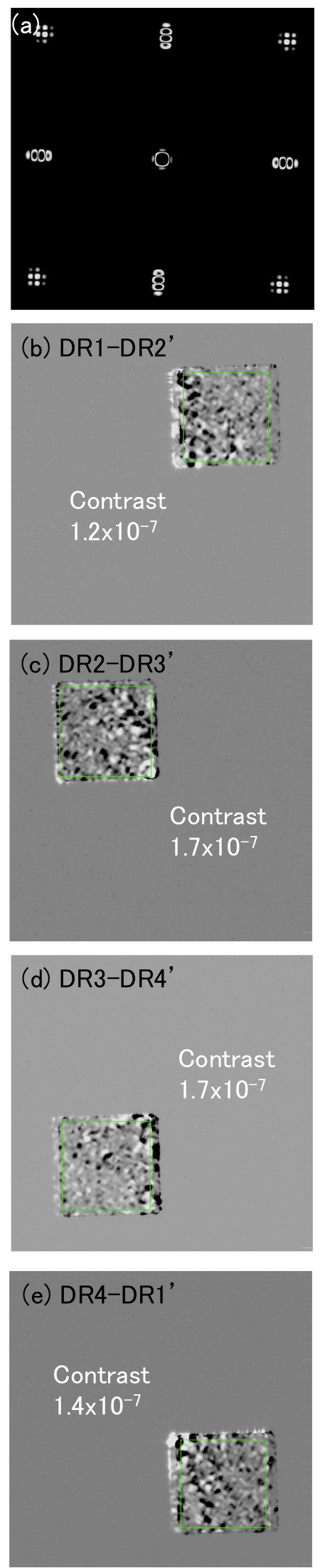}
                                             \end{center}
                                            \caption{
                                            Rotated mask subtraction 2: the way to subtract the images after from before mask rotation. 
                                            Panel (a):  shifting the core image before mask rotation to fit the centers of the core before and after mask rotation. 
                                            Panel (b), (c), (d) and (e):  images of the results of rotated mask subtraction 2 (DR1-DR2', DR2-DR3', DR3-DR4', DR4-DR1'). }
                                             \label{fig15}
                                           \end{figure}

We simulated the PSFs had only a mask shape error. 
We considered a simplified case of across-the-board increase in the line width of mask pattern where block the light.  
Fig.\ref{Fig1-5} shows simulated PSFs from across-the-board 5,10,15,20$\mu$m increases in the line width of the free-standing mask, respectively.  
In the case of across-the-board 20$\mu$m increase in the line width, the smallest holes, where pass through the light, in the mask design are closed. 
When we checked holes of the free-standing mask by using a digital microscope (VHX-900, KEYENCE), we found holes in the position of the holes expected from the mask design. 
Thus, the case of across-the-board 20$\mu$m increase is worse than in reality and extreme case. 
We ignored the case of across-the-board more than 20$\mu$m increase. 
Even in the case of across-the-board 20$\mu$m increase in the line width, the contrast of the dark region is not worse than $10^{-7}$. 
Therefore it was found that the factor of speckle is not only the mask shape error.  
This result of the numerical simulations of some mask shape errors is consistent with the result of the Rotated mask subtraction 1.

%
%
%

%

\chapter{Discussion}
         \section{Merits of the binary-shaped pupil mask coronagraph for an actual observation}

 We carried out a  series of studies of binary-pupil mask in order to address the critical issues on a binary-shaped pupil mask coronagraph on telescopes for direct observation of exoplanets. 
 As mentioned in the Introduction chapter, the critical issues are follows.
\begin{enumerate}
 \item The WFE caused by imperfections in the optics is an important limiting factor in the contrast of a coronagraph in space telescopes. 
 \item An actual observation is multi-color/broadband observation, not monochromatic observation. 
 \item An existing checker-board mask with a glass substrate can be the problem 
 that light loss by transmission, ghosting from residual reflectance and a slightly different refractive index for each wavelength.  
 \end{enumerate}
 We discussed three experiments on the issues. 


             \begin{enumerate}
           
                   \item As a result of using PSF subtraction, we improved the contrast by around two orders of magnitude from the raw contrast at the He-Ne laser wavelength. 
                   We demonstrated that the contrast of binary-shaped pupil mask coronagraph with PSF subtraction 
                   to achieve the contrast of $\sim$$10^{-9}$ required to observe giant planets directly.
                   The result confirmed that PSF subtraction of stable images is potentially beneficial for improving contrast in the laboratory. 
                   If enough stability and improvement in contrast thanks to PSF subtraction can be expected during coronagraph observations with SPICA or other telescopes, 
                   the requirements for the raw contrast can be significantly relaxed. 
                   Thus, the requirements for the static WFE can also be relaxed in a stable environment. 
                   This therefore can provide a significant advantage for the development of telescopes.



          　　　 \item The first results of our demonstration using multi-color/broadband sources have shown that 
                    the binary-shaped pupil mask coronagraph produces significant improvement in contrast at various wavelength bands, 
                    compared with non-coronagraphic optics. 
                    This result provide a significant advantage in installing a binary-shaped pupil mask coronagraph in a telescope, 
                    because it is necessary to make observations over a wavelength band, and it would be beneficial to make observations using multiple bands for an actual observation.


                      \item The free-standing mask was demonstrated a capacity to improve high-contrast of $1.0\times10^{-7}$ 
                      achieved for the raw coronagraphic image by areal averaging of all of the observed dark regions. 
                      Because a free-standing masks don't have wavelength dependence such a substrate mask, 
                      it can be use in the infrared region.  
                      For instance, it also can be installed in the next-generation infrared telescope SPICA. 
                      Nevertheless mask pattern has to be changed from checker-board design to new mask pattern 
                      considering the pupil shape of the telescope with a secondary mirror and spiders\citep{Enya2010b}.

                       
  \end{enumerate}

          \section{PSF subtraction at different wavelengths}
          
         Since we had both multi-color/broadband results, we applied the PSF subtraction at different wavelengths. 
         In principle, PSF subtraction of images at different wavelength cannot completely remove WFE, 
         even static WFE, because WFE has wavelength dependence. 
         On the other hand, if wavelength difference is small, noise pattern is expected to be similar. 
         So we consider PSF subtraction has practical effectiveness for improving contrast (e.g., \cite{Biller}). 
         We used images at 650nm and 750nm which had only a small amount of ghosting. 
         Because the image size is proportional to $\lambda/D$, 
         the image at 650nm was enlarged by 750/650 to make it the same size as the image at 750nm. 
         The positions were aligned by using a bright diffraction pattern outside the dark region as a guide. 
         The intensity was adjusted to the peak of the core image at 750nm. 
         We adjusted the size, position and intensity as described above and subtracted the image at 650nm from image at 750nm. 
         As a result, the areal mean of the dark region had a positive value. 
         In other words, the dark region in 750nm was brighter than the dark region in 650nm. 
         We believe this is because the ghost image is more pronounced at longer wavelengths. 
         The contrast between the peak of the core at 750nm and the areal mean of the dark region after PSF subtraction was 
         7.8$\times 10^{-7}$. 
         The contrast was a little better than 1.1$\times 10^{-6}$ which is the raw contrast of 750nm, 
         though the improvement in contrast was not as much as that obtained in the PSF subtraction experiment with the He-Ne laser. 
         Fig.\ref{fig7} shows no significant elongated pattern expected from difference in $\Delta\lambda/\lambda_c$ at 650nm and 750nm. 
         If we use images at longer wavelengths, the ghosting increases. 
         Thus, there is a need for the installation of a mirror optics system to improve the effects of ghost images.


　　　　　　　　　　　\begin{figure}[tbp]
                       \begin{center}
                          \includegraphics*[width=140mm]{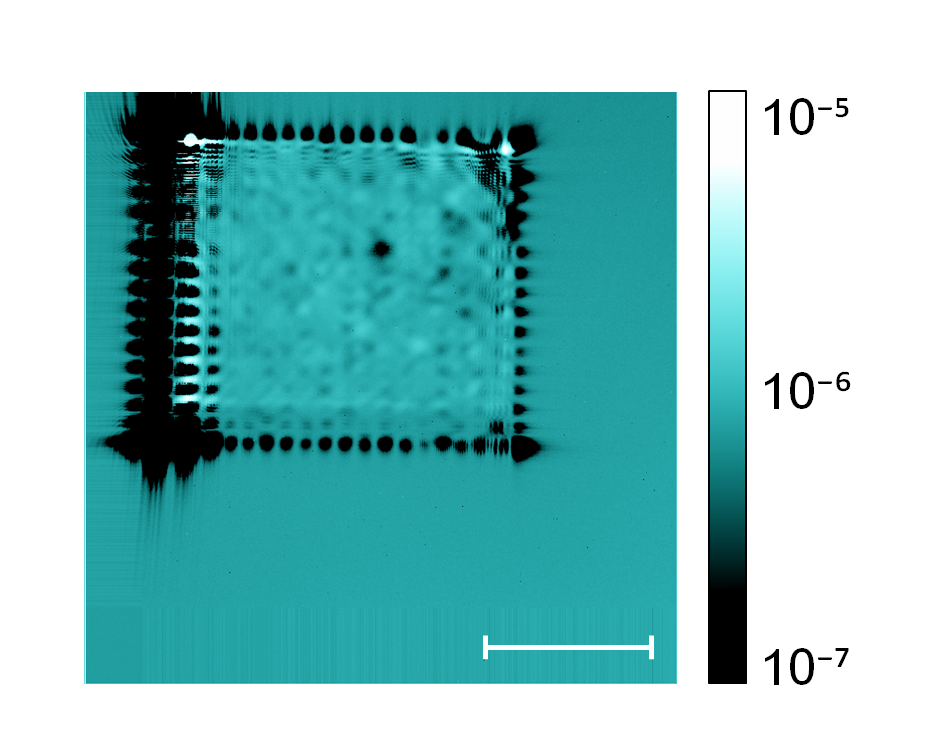}
                       \end{center}
                       \caption{An image of the result of PSF subtraction using two raw coronagraphic images at 650nm and 750nm. 
                       The scale bar is $10\lambda/D$. 
                       }
                       \label{Fig7}
                      \end{figure}

%
%
%


         %
         %
         \section{Comparison with other experiment }



  The development race was being conducted around the world (e.g., review in \cite{Guyon2006}). 
  Demonstration experiments are necessary because theory is ahead of experiments and high accuracy is required to make a theory more concrete. 
  In this section, we describe the world coronagraph experiments for space-telescopes. 
  There are two ways of space-coronagraphs;  one is focusing on direct observation of light reflected by the exoplanet at visible wavelengths and the other is focusing on direct observation of thermal emission from the exoplanets at infrared wavelengths. \\  

  
  
  
   The Terrestrial Planet Finder Coronagraph (TPF-C) is a coronagraph space mission capable of detecting and characterizing Earth-like planets and planetary systems at visible wavelengths around nearby stars. 
   TPF-C will use spectroscopy to measure key properties of exoplanets including the presence of atmospheric water or oxygen, powerful signatures in the search for habitable worlds. 
   A band-limited (BL) coronagraph was selected for the TPF-C Flight Baseline (FB1) architecture as it was the most mature technique, having been demonstrated in the laboratory to perform at levels needed for detecting Earths, $5.2\times10^{-10}$ at 4 $\lambda$/D for 760-840nm (10\% band) in natural unpolarized light, thus verifying the fundamental physics and establishing its feasibility \citep{Kern2008, Trauger}. 
   Other space missions with similar vision has been proposed, which aim direct observation of Earth-like exoplanets using a off-axis  space telescopes and various kind of coronagraph. \\
   
  
  The James Webb Space Telescope (JWST) is a space mission, which has a 6.5 m segmented aperture with four types of instruments, and will be launched in 2014. 
  Several coronagraphs will be installed in JWST: 
  One of the objectives of MIRI, the Mid-InfraRed Instrument of the JWST (5-28 μm), is the direct detection of Extrasolar Giant Planets (EGPs) around young stars \citep{Cavarroc}. 
  Four coronagraphs are installed in MIRI. One is a Lyot coronagraph operating at 23 μm optimized for the study of circumstellar disks, 
  the three others are monochromatic FQPMs operating at 10.65 μm, 11.4 μm, and 15.5 μm.  
  They are associated with narrowband filters (the spectral bandpass is 5\% for the FQPMs and 20\% for the Lyot coronagraph). 
  They are optimized for the detection and characterization of Jupiter-like planets \citep{Boccaletti, Baudoz2006}. 
  It also includes the NIRCam coronagraph between 2.1 μm and 4.6 μm \citep{Krist} and the Non-Redundant Masking on the Fine Guidance Sensor Tunable Filter Imager (FGS-TFI/NRM) between 3.8 μm and 5.5 μm \citep{Sivaramakrishnan}. \\

   SPICA (2018 launch planned)  is an astronomical mission optimized for mid-infrared and far-infrared astronomy with a 3m class on-axis telescope cooled to $<$6K \citep{Nakagawa}.
The primary target of the SPICA coronagraph is a self-luminous Jovian extra-planets around 1-5 Gyr old G-M type stars. 
Binary shaped pupil mask coronagraph has potential for SPICA coronagraph  \citep{Enya2010a}. \\

 Next, we refer to laboratory testing for space-telescopes.    
  Several concepts exist, many are being prototyped across the world, and some have already demonstrated. \\
  
 Phase Mask laboratory testing: 
 Monochromatic device performance has already been demonstrated and the manufacturing procedures are well-under control since their development. 
 Among them, the Annular Groove Phase Mask \citep{Mawet, Foo}, the Four-Quadrant Phase Mask (FQPM) \citep{Carlotti} , and the Eight-Octant Phase Mask (EOPM) \citep{Murakami} are quite promising. 
 The multistage four-quadrant phase mask (MFQPM) reduces the stellar flux over a  spectral range and it is a very good candidate to be associated with a spectrometer for future exoplanet imaging instruments in ground- and space-based observatories \citep{Baudoz2008}. The coronagraph gives an average transmission between $7\times10^{-6}$ and $4\times10^{-5}$  at each wave- length over a 20\% bandwidth (660-800 nm). \citep{Galicher}. \\

   PIAA laboratory testing: The laboratory experiment achieved a $2.27\times10^{-7}$ raw contrast 
   between 1.65 $\lambda$/D (inner working angle of the coronagraph configuration tested) and 4.4 $\lambda$/D (outer working angle) \citep{Guyon2010}. 
   The NASA Ames Research Center PIAA coronagraph laboratory is a highly flexible testbed operating in air \citep{Belikov2009}. 
   It is dedicated to PIAA technologies and is ideally suited to rapidly develop and validate new technologies and algorithms. 
   It uses MEMS-type deformable mirrors for wavefront control. 
   The NASA JPL High Contrast Imaging Testbed (HCIT) is a high stability vacuum testbed facility for coronagraphs. 
   PIAA is one of the coronagraph techniques tested in this lab, which provides the stable vacuum environment ultimately required to validate PIAA for flight \citep{Kern} .  \\

  Pupil-plane Mask Coronagraph laboratory testing: \cite{Belikov} achieved the contrast of $4\times 10^{-8}$ using visible laser, and $\sim10^{-7}$ contrast using broadband light source with speckle nulling in a small area from 4$\lambda/D$ to 9$\lambda/D$.  \\


There are many of the various kinds of coronagraph experiments which have a different purpose and a different demonstration performance. 
     
The common characteristic of TPF-C and other missions using visible coronagraph,
JWST, and SPICA is a space-coronagraph.    
TPF-C and visible coronagraph missions are aimed at direct observation of light reflected by the exoplanet (ultimately, the Earth-like planet) at visible wavelengths. 
On the other hand, JWST and SPICA are aimed at direct observation of thermal emission from the exoplanet at infrared wavelengths. 
Larger entrance aperture and earlier launch are big advantage of JWST.
For SPICA,  a binary-pupil mask coronagraph, which is used in this study and candidate of coronagraph, is suitable for spectroscopic observation in principle, 
because it is effective over a  range of wavelengths.
Meanwhile, the FQPM and the Lyot coronagraph, which is candidate for JWST coronagraph, has wavelength dependence. 
In recent years, however, MFQPMs and apodized-pupil Lyot coronagraphs involving multistage configuration can reduce the stellar flux at several wavelengths. 

   The common thing between the study on HCIT and our study is using a vacuum chamber. 
   The range of different coronagraphs such as the binary-pupil mask coronagraph, PIAA and Lyot coronagraph is studied on the HCIT. 
   Multi-color experiments with speckle nulling were conducted on HCIT. 
   On the other hand, in this study, we used only the binary-pupil mask coronagraphs. We conducted the multi-color/broadband experiments without speckle nulling. 

         It is beneficial to compare this work with other experiment using a binary-shaped pupil mask coronagraph. 
         \cite{Belikov} achieved the contrast of $4\times 10^{-8}$ using visible laser, 
         and $\sim10^{-7}$ contrast using broadband light source 
         with speckle nulling in a small area from 4$\lambda/D$ to 9$\lambda/D$. 
         On the other hand, our experiments in this paper were performed without speckle nulling. 
         The works shown in this paper resulted  the contrast of $2.3 \times 10^{-7}$ using He-Ne laser, 
         which is just close to the limit of the mask design, and $\sim10^{-6}$ contrast using broadband, 
         in a larger area, from 3$\lambda/D$ to 30$\lambda/D$. 
         Therefore, these works are complementary to each other. 
         It is a possible future work of us to introduce speckle nulling 
         in order to reach higher contrast \citep{Kotani}. 
         In such case, the design of our free-standing mask can be useful because the contrast 
         produced by this mask is designed to be $10^{-10}$. 
         
%
         
           \section{Future works}       


 There are two ways of developing this work. 
One is focusing on direct observation of light reflected by the exoplanet using a space coronagraph at visible wavelengths. 
This means that the coronagraph must achieve the contrast of $10^{-10}$ at visible wavelengths. 
It was found that a major limiting factor for contrast is WFE. 
If proper wavefront control is used in this binary-shaped pupil mask coronagraph experiment, the improving contrast is expected. 
Compared to other previous study (i.e., HCIT), the raw contrast is very high in this study. 
It is interesting that pursuit of higher contrast with the vacuum chamber and wavefront control. 
The other is focusing on direct observation of thermal emission from the exoplanet using a space coronagraph at infrared wavelengths. This means that the coronagraph must achieve the contrast of $10^{-6}$ at infrared wavelengths. 
To achieve this, we must reduce thermal noise. It requires application of a cryogenic vacuum chamber to reduce thermal noise. 
The glass lenses cannot be used and mirror optics are appropriate for infrared experiments.
We need infrared sources and a infrared detector. 
A glass substrate mask cannot be used. 
A free-standing mask used for this study is a very promising coronagraph mask because of regardless of substrate property.

 \chapter{Summary and Conclusion}
  Direct detection and spectroscopy of exoplanets is essential for understanding 
   how planetary systems were born, how they evolve, and, ultimately, 
   for finding biological signatures on these planets. 
   The enormous contrast in luminosity between the central star and a planet presents the primary difficulty 
   in the direct observation of exoplanets. 
   For example, if the solar system is observed from a distance, 
   the expected contrast between the central star and the planet  
   is $\sim10^{-10}$ in the visible light region and $\sim10^{-6}$ in the mid-infrared region. 
   Therefore, a stellar coronagraphs which can improve the contrast between the star and the planet have to be developed. 
   Of the various kinds of coronagraphs, we focused on a binary-shaped pupil mask coronagraph. 
   The reasons for using this coronagraph are that it is robust against pointing errors, 
can make observations over a wide range of wavelengths in principal and is relatively simple.  
   Also, the adoption of a binary-shaped pupil mask coronagraph for SPICA is considered.  
   We conducted a number of coronagraph experiments using a vacuum chamber and a checker-board mask, a kind of binary-shaped pupil mask, 
   without active wavefront control. 
   This study is unique and important with respect to not only the tasks necessary to make the coronagraph fit for practical use, 
   but also the verification tests for more real coronagraphic observations. 
   Three kinds of experiment in this study are described below. 
 \begin{enumerate}
\item In space telescopes, the WFE caused by imperfections in the optics is an important limiting factor in the contrast of a coronagraph.  
 Subtraction of PSF is beneficial in that it removes any static WFE, and achieves  a higher contrast than the raw contrast of coronagraph.   
PSF subtraction is available in direct observations of exoplanets using space telescopes, which helps to improve high-contrast observations.  
We evaluated how much the PSF subtraction contributes to the high contrast observation by subtracting the images obtained through the coronagraph. 
We improved the temperature stability by installing the coronagraph optics in a vacuum chamber, controlling the temperature of the optical bench, and covering the vacuum chamber with thermal insulation layers. 
A contrast of 2.3$\times 10^{-7}$ was obtained for the raw coronagraphic image and a contrast of 1.3$\times 10^{-9}$ was achieved after PSF subtraction with a He-Ne laser at 632.8nm wavelength. 
Thus, the contrast was improved by around two orders of magnitude from the raw contrast by subtracting the PSF. 
\item A He-Ne laser was employed as the light source in the previous experiments. 
Meanwhile, the binary-shaped pupil mask coronagraph should work at all wavelengths in principle.   
 For an actual observation, it is necessary to make observations over a wavelength band, and it would be beneficial to make observations using multiple bands.    
We carried out multi-color/broadband experiments using SLEDs 
with center wavelengths of 650nm, 750nm, 800nm and 850nm in order to demonstrate a more realistic observation. 
 We achieved contrasts of 3.1$\times 10^{-7}$, 1.1$\times 10^{-6}$, 1.6$\times 10^{-6}$ and 2.5$\times 10^{-6}$ 
at the bands of 650nm, 750nm, 800nm and 850nm, respectively. 
The results show that contrast within each of the wavelength bands was significantly improved compared with non-coronagraphic optics. 
\item However, an existing checker-board mask with a glass substrate has the problems 
of light loss by transmission, ghosting from residual reflectance and a slightly different refractive index for each wavelength.  
Therefore, we developed a new free-standing mask with sheet metal without substrate. 
The free-standing mask is available in the infrared observation which has a grate advantage over the visible light observation in the contrast between the star and the planet. 
As the result of He-Ne laser experiment with the free-standing mask, 
we achieved contrasts of $1.2\times10^{-7}$、$9.5\times10^{-8}$、$9.6\times10^{-8}$、$9.6\times10^{-8}$ at the DR1, DR2, DR3 and DR4, respectively. 
Speckles are the major limiting factor. 
Thus, the free-standing mask demonstrated about the same ability to improve the contrast significantly as the substrate mask. 
\end{enumerate}

We demonstrated PSF subtraction is potentially beneficial for improving contrast of a binary-shaped pupil mask 
coronagraph, 
this coronagraph produces a significant improvement in contrast 
with multi-color/broadband light sources, 
and the new free-standing mask for practical use provides superior performance of improving contrast. 
We performed the tasks necessary to make the coronagraph fit for practical use. 
In conclusion, we carried out verification tests for more real coronagraphic observations. 
This study developed the existing study of a checker-board mask coronagraph and overcame difficulties of practical observation by the satellite. 
Consequently, this study suggests that a binary-shaped pupil mask coronagraph can be applied to coronagraphic observation 
by SPICA and by other telescopes.

\bigskip

\section*{Acknowledgements} 
I am deeply indebted to Professor Matsuhara that he gave me great opportunity to study infrared astronomy at ISAS/JAXA.
I thank Professor Nakagawa for his invaluable advice and recommendations on the experiment and this paper. 
I am grateful to Assistant Professor Enya for his enormous support and insightful comments. 
His comments and suggestions were of inestimable value for this study. 
I thank T. Kotani since his support for wavefront simulation. 
Discussions with L. Abe were very fruitful for this study.
I am grateful to the pioneers of the checker-board pupil mask coronagraph, 
especially R. J. Vanderbei. 
This research was partially supported by Grant-in-Aid for JSPS Fellows.
The pupil mask was supported by the Nano-Processing Facility of the Advanced Industrial Science and Technology. 
The Free-standing mask was made with Howa Sangyo Co., Ltd and Photo Precision Co., Ltd. 
I thank SIGMA KOKI CO., LTD. for providing BMAR data. 

I would like to express my thanks to the members of the department of infrared astrophysics in ISAS/JAXA for their comments made enormous contribution to my work. 
I would also like to express my gratitude to my family for their continuing support and warm encouragements.
       


\newpage

\section*{Appendix}

This section shows manufacture process of 10mm size checkerboard masks;
mask on substrate of Silicon, Germanium, BK7, and 
free standing mask of Cu, Ni.
In these masks, one free standing mask made of Cu was used in 
experiments in this thesis.
Si, Ge masks are made for mid-infrared tests in future, and BK7 masks are
comparison reference (they can be conveniently tested with 
visible light).
Free standing masks are also hopeful for infrared coronagraph.
All these masks have same size and same design to realize
systematic tests and comparison for infrared coronagraph 
in future.

\vspace{5mm}

(The mask with glass substrate evaluated in this thesis was
2mm size mask, so different one from BK7 masks described in this
appendix).

\vspace{5mm}

More detail process of the manufacture is shown below. 
First, a photomask was manufactured. 
This photomask was used for all 10mm size masks.
Next, manufacture process of masks on Si, Ge, BK7 substrate,
and then the manufacture process of free standing masks of Cu, Ni
is shown.

\vspace{5mm}

Whilst main purpose of these masks is tests for mid-infrared 
coronagraph like SPICA, it is also interesting that
some of these masks have potential fur use in near infrared 
and visible wavelength region.

  
\subsection*{Fabrication process of a photomask for microstructure patterning}
\begin{enumerate}
\item 4-inch (101.6mm) squared-shape glass is used as a substrate
\item Spattering of $Cr+Cr_{2}O_{3}$ (0.1 $\mu$m thickness) on the substrate
     (Spattering is a coating process with a fine layer of metal, 
     by driven by momentum exchange between the ions and atoms in the materials by collisions)
\item Spin coat resist on $Cr_{2}O_{3}$ with a thickness of 0.5 $\mu$m 
     (Spin-coat: a procedure used to make uniform thin layer to flat substrates by 
     rotating the substrate at high speed in order to spread the fluid by centrifugal force)
\item Exposure: drawing of the microstructure patterns with 412 nm laser. 
      Development: Remove the photoresist (a positive type)
\item Wet etching of the substrate in a acid etching solution in a thin polyvinyl container
\item Strip the resist with a remover
\end{enumerate}


\begin{figure}[tbp]                                            
 \begin{center}                                                
  \includegraphics*[width=70mm]{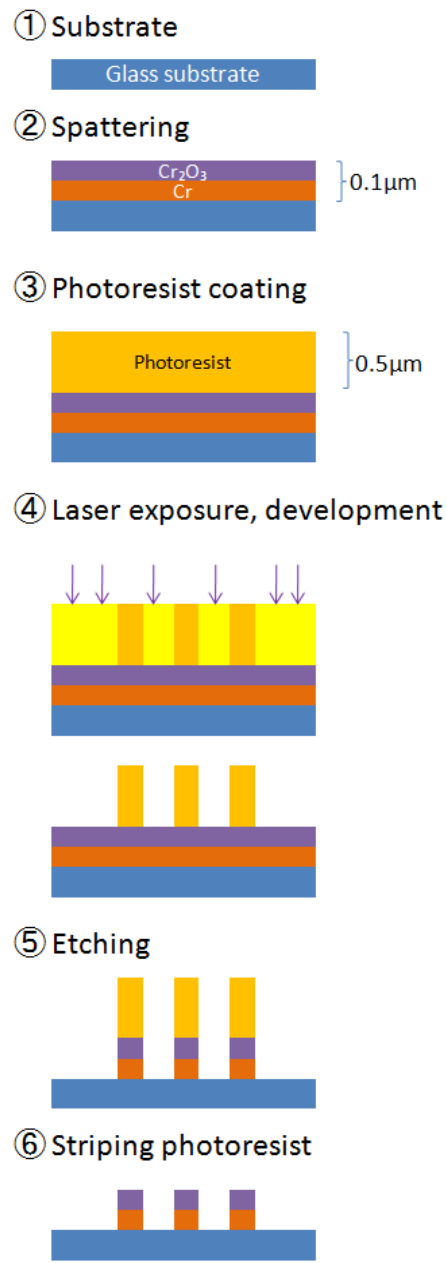}
 \end{center}
    \caption{Fabrication process of a photomask for microstructure patterning}
    \label{fig8}
\end{figure}


\subsection*{Fabrication process of a mask on a BK7 glass, Silicon, and Germanium substrate}
\begin{enumerate}
\item $\phi$ 30mm glass substrate, 50mm-squared Silicon and Germanium substrate are used
\item EB vapor deposition of Aluminium on the substrate with a thickness of 
      1000 \AA, 2000 \AA, 4000 \AA, 8000 \AA, and 1.6 $\mu$m (EB vapor deposition is a 
      type of vapor deposition in which heating is done by Electron Beam)
\item Spin coat of photoresist on Al with a thickness of 1 $\mu$m
\item Exposure: transfer the patterns from the photomask to the photoresist by UV light (365nm). 
      Development: Remove the photoresist (a positive type)
\item Wet etching of the substrate in a acid etching solution in the thin polyvinyl container
\item Strip the photoresist with acetone dip
\end{enumerate}


\begin{figure}[tbp]                                             
   \begin{center}
   \includegraphics*[width=70mm]{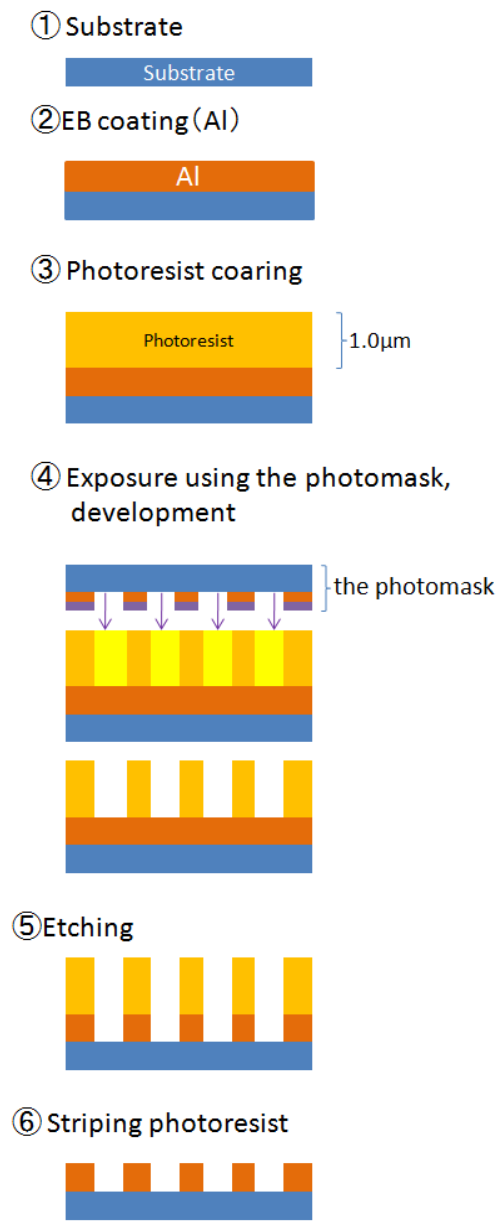}
   \end{center}
   \caption{Fabrication process of a mask on a BK7 glass substrate.
    Same processes were applied for other substrates.}
   \label{fig4}
\end{figure}


\subsection*{Fabrication process of a Cu and Ni free-standing mask}
\begin{enumerate}
\item Spattering of $Cr+Cr_{2}O_{3}$ (0.1 $\mu$m thickness) on 4-inch (101.6mm) squared-shape 
      glass substrate, then attach a substrate on which resist is spin-coated (commercially 
      available one with a good surface accuracy) on the glass substrate. Remove the resist 
      and $Cr+Cr_{2}O_{3}$ in order to use only the glass substrate
\item Spin-coating mold releasing agent with a thickness of 1 $\mu$m on the surface
\item EB vapor deposition of Cu (0.5 $\mu$m thickness) on the release agent
\item Spin-coat resist (with a goal of 10 $\mu$m thickness) for plating on the Cu substrate
\item Exposure: transfer the patterns from the photomask to the photoresist by UV light (365nm). 
      Development: Remove the photoresist (only the illuminated part is resolved, a positive type) 
\item Electrolytic plating of Cu with a goal thickness of 2, 5, 10, 20 $\mu$m. 0.5 $\mu$m thickness 
      Cu is used as a seed layer for plating seed layer (Electrolytic plating is a plating 
      process in which metal ions in a solution are moved by an electric field for coating. 
      The process uses direct electrical current to reduce positive ions of a desired material 
      from a solution and coat a conductive material (here it is a seed layer) 
      with a thin layer of the material).
\item Laminate the dry film resist (100 $\mu$m thickness) in order for plating the outer 
      holder region) after microstructure plating
\item Exposure: transfer the patterns from the photomask to the photoresist by UV light (365nm). 
      Development: Remove the resist. Only the illuminated part of the dry film resist remains (a negative type)
\item Electrolytic plating of Cu with a goal thickness of 100 $\mu$m. 
      0.5 $\mu$m thickness Cu is used as a seed layer for seed layer
\item Strip the dry film resist with acetone dip
\item Strip the resist with acetone dip
\item Etching the Cu seed layer (0.5 $mu$m thickness). Wet etching of the substrate in 
      an acid etching solution in a thin polyvinyl container. Not only a seed layer but 
      also 0.5 $\mu$m Cu layer is etched and removed   
\item Strip the release agent in the number 2 with the acetone dip
\item Rinse the substrate with IPA (isopropyl alcohol) then natural drying
\end{enumerate}

\begin{figure}[tbp]
    \begin{center}
    \includegraphics*[width=150mm]{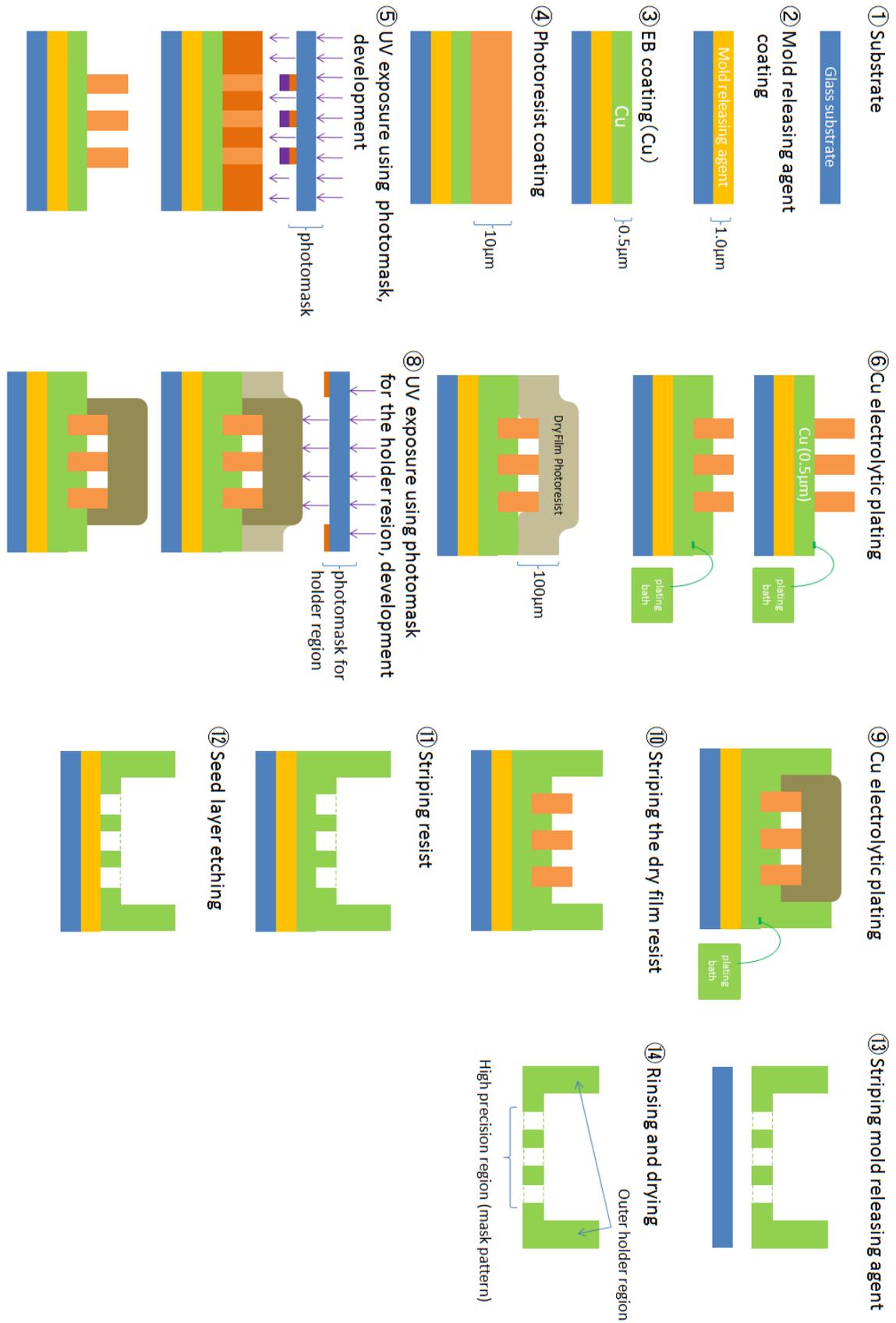}
    \end{center}
    \caption{Fabrication process of a Cu free-standing mask.
       Same process was applied for Ni free-standing mask.}
    \label{fig9}                                            
\end{figure}


%
%



\begin{thebibliography}{99}


\bibitem[Abe et al.(2001)]{Abe}
Abe, L., Vakili, F., \& Boccaletti, A. 2001, A\&A, 374, 1161

\bibitem[Aime(2001)]{Aime2001}	
Aime, C., Soummer, R., \& Ferrari, A. 2001, A\&A, 379, 697

\bibitem[Aime(2005)]{Aime}	
Aime, C. 2005, A\&A, 434, 785

\bibitem[Aime \& Soummer (2004)]{AimeSoummer}
Aime, C., \& Soummer, R. 2004, Proc. SPIE, 5490, 456

\bibitem[Baudoz et al.(2006)]{Baudoz2006}	
Baudoz, P., Boccaletti, A., Riaud, P., Cavarroc, C., et al. 2006, PASP, 118, 765

\bibitem[Baudoz et al.(2008)]{Baudoz2008}	
Baudoz, P., Galicher, R., Baudrand, J., \& Boccaletti, A., 2008, SPIE, 7015, 176

\bibitem[Beaulieu et al.(2006)]{Beaulieu}
Beaulieu, J.-P., Bennett, D. P., Fouque', P., et al., 2006, Nature, 439, 437

\bibitem[Belikov et al.(2006)]{Belikov}
Belikov, R., Give\' on, A., Trauger, J. T., et al. 2006, SPIE, 6265, 18

\bibitem[Belikov et al.(2009)]{Belikov2009}
Belikov, R., Pluzhnik, E., Connelley, M. S., et al. 2009, SPIE, 7440, 17

\bibitem[Bennett et al.(2008)]{Bennett}
Bennett, D. P., Bond, I. A., Udalski, A., et al., 2008, ApJ, 684, 663

\bibitem[Biller et al.(2009)]{Biller}
Biller, B., Trauger, J., et al. 2009, PASP, 121, 881, 716

\bibitem[Boccaletti et al.(2005)]{Boccaletti}
Boccaletti, A., Baudoz, P., Baudrand, J., Reess, J. M., Rouan, D., 2005, ASR, 36, 1099

\bibitem[Burrows et al.(2003)]{Burrows}
Burrows, A, Sudarsky, D, Lunine, J. I.. 2003, ApJ, 596, 587

\bibitem[Carlotti et al.(2009)]{Carlotti}
Carlotti, A., Ricort, G., \& Aime, C., 2009, A\&A, 504, 663

\bibitem[Carlotti et al.(2011)]{Carlotti2011}
Carlotti, A., Vanderbei, R., \& Kasdin, N. J., 2011, arXiv:1108.4050

\bibitem[Cavarroc et al.(2008)]{Cavarroc}
Cavarroc, C., Amiaux, J., Baudoz, P., Boccaletti, A., et al. 2008, SPIE, 7010, 29

\bibitem[Charbonneau et al.(2000)]{Charbonneau}
Charbonneau, D., Brown, T. M., Latham, D. W., \& Mayor, M. 2000, ApJ, 529, 45

\bibitem[Codona \& Angel(2004)]{Codona}
Codona, J. L., \& Angel, R. 2004, ApJ, 604, L117

\bibitem[Deming et al.(2005)]{Deming}
Deming, D., Seager, S., Richardson, L. J., \& Harrington, J. 2005, Nature, 434, 740

\bibitem[Enya et al.(2007)]{Enya2007a}
Enya, K., Tanaka, S., Abe, L., \& Nakagawa, T. 2007, A\&A, 461, 783


\bibitem[Enya et al.(2010)]{Enya2010a}
Enya, K., \& SPICA working group, 2010, ASR, 45, 979

\bibitem[Enya \& Abe(2010)]{Enya2010b}
Enya, K., \& Abe, L. 2010, PASJ, 62, 1407

\bibitem[Enya et al.(2011)]{Enya2011}
Enya, K., Abe, L., Takeuchi, S., Kotani, T., \& Yamamuro, T., 2011, Proc. of SPIE, in press

\bibitem[Foo et al.(2005)]{Foo}
Foo, G., Palacios, D. M., \& Swartzlander, G. A., Jr., 2005, Opt. Lett., 30, 3308

\bibitem[Galicher et al.(2011)]{Galicher}
Galicher, R., Baudoz, P., Baudrand, J., 2011, A\&A, 530, 43 

\bibitem[Green et al.(2004)]{Green}
Green, J. J., Shaklan, S. B., Vanderbei, R. J., \& Kasdin, N. J. 2004, SPIE, 5487, 1358

\bibitem[Gonsalves \& Nisenson(2003)]{Gonsalves}
Gonsalves, R., \& Nisenson, P. 2003, PASP, 115, 706


\bibitem[Guyon(2003)]{Guyon2003}
Guyon, O. 2003, A\&A, 404, 379

\bibitem[Guyon et al.(2005)]{Guyon2005}	
Guyon, O., Pluzhnik, E., Galicher, R., et al., 2005, ApJ, 622, 744

\bibitem[Guyon et al.(2006)]{Guyon2006}
Guyon, O., Pluzhnik, E. A., Kuchner, M. J., Collins, B., \& Ridgway, S. T., 2006, ApJ, 167, 81

\bibitem[Guyon et al.(2010)]{Guyon2010}
Guyon, O., Pluzhnik, E., Martinache, F., et al.2010, PASP, 122, 71

\bibitem[Guyon \& Roddier(2000)]{Guyon2000}
Guyon, O., \& Roddier, F. J. 2000, Proc. SPIE, 4006, 377

\bibitem[Haze et al.(2009)]{Haze}
Haze, K., Enya, K., Abe, L., et al. 2009, ASR, 43, 181


\bibitem[Jacquinot \& Roizen-Dossier(1964)]{Jacquinot}
Jacquinot, P., \& Roizen-Dossier, B. 1964, Prog. Opt., 3, 29

\bibitem[Kalas et al.(2008)]{Kalas2008}
Kalas, P., Graham, J. R., Chiang, E., et al. 2008, Science, 322,1345

\bibitem[Kasdin et al.(2003)]{Kasdin}
Kasdin, N. J., Vanderbei, R. J., Spergel, D. N., \& Littman, M. G. 2003, ApJ, 582, 1147

\bibitem[Kasdin et al.(2005a)]{Kasdin2005a}
Kasdin, N. J., Belikov, R., Beall, J., Vanderbei, R. J., et al. 2005, Proc. of SPIE, 5905, 128

\bibitem[Kasdin et al.(2005b)]{Kasdin2005b}
Kasdin, N. J., Vanderbei, R. J., Littman, M. G., \& Spergel, D. N., 2005, Appl. Opt., 44, 1117

\bibitem[Kern et al.(2009)]{Kern}
Kern, B., Belikov, R., Give'On, A., Guyon, O., et al. 2009, SPIE, 7440,15

\bibitem[Kern et al.(2008)]{Kern2008}
Kern, K \& Trauger, 2008,  JPL Publication D-60951

\bibitem[Kotani et al.(2010)]{Kotani}
Kotani, T., Enya, K., Nakagawa, T., et. al., 2010, Proc. ASP Conference Series, 430, 477

\bibitem[Krist  et al.(2009)]{Krist}
Krist, J. E., Balasubramanian, K., Beichman, C. A., et. al. 2009, SPIE, 7440, 28

\bibitem[Kuchner et al. (2005)]{Kuchner}
Kuchner, M. J., Crepp, J., \& Ge, J. 2005, ApJ, 628, 466

\bibitem[Kuchner \& Traub (2002)]{KuchnerTraub}
Kuchner, M. J., \& Traub, W. A. 2002, ApJ, 570, 900

\bibitem[Lyot(1939)]{Lyot}
Lyot, B. 1939, MNRAS, 99, 580

\bibitem[Marois et al.(2008)]{Marois2008}
Marois, C., Macintosh, B., Barman, T., et al. 2008, Science, 322, 1348

\bibitem[Martinache et al.(2006)]{Martinache}
Martinache, F., Guyon, O., Pluzhnik, E. A., et al. 2006, ApJ, 639, 1129


\bibitem[Mayor  \& Queloz(1995)]{Mayor}
 Mayor, M., \& Queloz, D., 1995, Science, 378, 6555

\bibitem[Mawet et al.(2005)]{Mawet}	
Mawet, D., Riaud, P., Absil, O., \& Surdej, J., 2005, ApJ, 633, 1191

\bibitem[Murakami et al.(2008)]{Murakami}	
Murakami, N., Uemura, R,. Baba, N., et al., 2008, PASP, 120, 1112

\bibitem[Nisenson \& Papaliolios(2001)]{Nisenson}	
Nisenson, P., \& Papaliolios, C. 2001, ApJ, 548, L201


\bibitem[Nakagawa et al.(2010)]{Nakagawa}
Nakagawa, T., \& SPICA team, 2010, SPIE, 7731, 18

\bibitem[Palacios(2005)]{Palacios}
Palacios, D. M. 2005, Proc. SPIE, 5905, 196

\bibitem[Pluzhnik et al.(2006)]{Pluzhnik}
Pluzhnik, E. A., Guyon, O., Ridgway, S. T., et al., 2006, ApJ, 644, 1246


\bibitem[Roddier \& Roddier(1997)]{Roddier}
Roddier \& Roddier, 1997, PASP, 109, 815

\bibitem[Rouan et al.(2000)]{Rouan}
Rouan, D., Riaud, P., Boccaletti, A., Clenet, Y., Labeyrie, A., 2000, PASP, 112, 1479

\bibitem[Sivaramakrishnan et al.(2009)]{Sivaramakrishnan}
Sivaramakrishnan, A., Tuthill, P., Martinache, F., et al.,  2009, Astro2010, 40

\bibitem[Slepian (1965)]{Slepian}
Slepian, D. 1965, J. Opt. Soc. Am., 55, 1110

\bibitem[Soummer et al. (2003a)]{Soummer2003a}
Soummer, R., Aime, C., \& Falloon, P. E. 2003a, A\&A, 397, 1161

\bibitem[Spergel(2001)]{Spergel}
Spergel, D. N. 2001, Astro-ph/0101142

\bibitem[Swain et al.(2009)]{Swain}
Swain, M. R., Tinetti, G., et al., 2009, ApJ, 704, 1616

\bibitem[Swartzlander(2006)]{Swartzlander}
Swartzlander, G. A., Jr. 2006, Opt. Lett., 31, 2042

\bibitem[Tanaka et al.(2006)]{Tanaka}
Tanaka, S., Enya, K., Abe, L., Nakagawa, T., \& Kataza, H.2006, PASJ, 58, 627

\bibitem[Tinetti et al.(2007)]{Tinetti}
Tinetti, G., Vidal-Madjar, A., 2007, Nature, 448, 169

\bibitem[Traub \& Jucks(2002)]{Traub}
Traub, W. A., \& Jucks, K. W., 2002, Astro-ph/0205369

\bibitem[Traub \& Vanderbei(2003)]{Traub2003}
Traub, W. A., \& Vanderbei, R. J. 2003, ApJ, 599, 695


\bibitem[Trauger \& Traub(2007)]{Trauger}
Trauger, J. T., \& Traub, W. A. 2007, Nature, 446, 771 

\bibitem[Vanderbei(1999)]{Vanderbei1999}
Vanderbei, R. J. 1999, Optimization methods \& software, 11, 485 

\bibitem[Vanderbei (2006)]{Vanderbei2006}
Vanderbei, R. J. 2006, ApJ, 636, 528

\bibitem[Vanderbei et al.(2003a)]{Vanderbei2003}
Vanderbei, R. J., Spergel, D. N., \& Kasdin, N. J. 2003a, ApJ, 590, 593

\bibitem[Vanderbei et al.(2003b)]{Vanderbei2003b}
Vanderbei, R. J., Spergel, D. N., \& Kasdin, N. J., 2003b, ApJ, 599, 686

\bibitem[Vanderbei et al.(2004)]{Vanderbei2004}
Vanderbei, R. J., Kasdin, N. J., \& Spergel, D. N. 2004, ApJ, 615, 555 

\bibitem[Vanderbei \& Traub(2005)]{Vanderbei2005}
Vanderbei, R. J., \& Traub, W. A. 2005, ApJ, 626, 1079 

\bibitem[Yang \& Kostinski(2004)]{Yang}
Yang, W., \& Kostinski, A. B. 2004, ApJ, 605, 892










\end{thebibliography}
\end{document}